\documentclass[pra, notitlepage, showpacs, floatfix, reprint, nobalancelastpage,citeautoscript,superscriptaddress]{revtex4-1}

\usepackage[T1]{fontenc}
\usepackage[latin9]{inputenc}
\usepackage{mathtools}

\usepackage{amsmath}
\usepackage{amssymb}
\usepackage{bbm}
\usepackage{braket}
\usepackage{xcolor}
\usepackage{pifont}
\usepackage[mathscr]{euscript}
\usepackage[shortlabels]{enumitem}

\usepackage{graphicx}
\usepackage{dcolumn}
\usepackage{bm}
\usepackage{subfig}
\usepackage{soul}

\newcommand{\equref}[1]{Eq.~(\ref{#1})}

\newcommand{\secref}[1]{Sec.~\ref{#1}}
\newcommand{\figref}[1]{Fig.~\ref{#1}}
\newcommand{\refcite}[1]{Ref.~\onlinecite{#1}}

\newcommand{\tableref}[1]{Table~\ref{#1}}

\renewcommand{\vec}[1]{\boldsymbol{#1}}

\usepackage{graphicx}
\usepackage[colorlinks=true]{hyperref}  
\hypersetup{
    bookmarks=true,         
    unicode=false,          
    pdftoolbar=true,        
    pdfmenubar=true,        
    pdffitwindow=false,     
    pdfstartview={FitH},    
    pdftitle={Generative models for sampling and phase transition indication in spin systems},    
    pdfauthor={Singh et al},     
    pdfsubject={},   
    pdfcreator={},   
    pdfproducer={}, 
    pdfkeywords={} {} {}, 
    pdfnewwindow=true,      
    colorlinks=true,       
    linkcolor=blue, 
    citecolor=blue,        
    filecolor=magenta,      
    urlcolor=blue           
} 

\begin{document}

\title{Generative models for sampling and phase transition indication in spin systems}

\author{Japneet Singh}
\affiliation{Department Of Electrical Engineering,
Indian Institute Of Technology Kanpur,
Kanpur, Uttar Pradesh, India}

\author{Vipul Arora}
\affiliation{Department Of Electrical Engineering,
Indian Institute Of Technology Kanpur,
Kanpur, Uttar Pradesh, India}

\author{Vinay Gupta}
\affiliation{Department Of Electrical Engineering,
Indian Institute Of Technology Kanpur,
Kanpur, Uttar Pradesh, India}

\author{Mathias S. Scheurer}
\affiliation{Department of Physics, Harvard University, Cambridge MA 02138, USA}

\date{\today}

\begin{abstract}
Recently, generative machine-learning models have gained popularity in physics, driven by the goal of improving the efficiency of Markov chain Monte Carlo techniques and of exploring their potential in capturing experimental data distributions. Motivated by their ability to generate images that look realistic to the human eye, we here study generative adversarial networks (GANs) as tools to learn the distribution of spin configurations and to generate samples, conditioned on external tuning parameters, such as temperature. We propose ways to efficiently represent the physical states, e.g., by exploiting symmetries, and to minimize the correlations between generated samples. We present a detailed evaluation of the various modifications, using the two-dimensional XY model as an example, and find considerable improvements in our proposed implicit generative model. It is also shown that the model can reliably generate samples in the vicinity of the phase transition, even when it has not been trained in the critical region.
On top of using the samples generated by the model to capture the phase transition via evaluation of observables, we show how the model itself can be employed as an unsupervised indicator of transitions, by constructing measures of the model's susceptibility to changes in tuning parameters.  
\end{abstract}

\maketitle

\section{\label{sec:level1}Introduction}
Generative models \cite{GenerativeModelsReview2015,GenerativeModelsforPhysicists,ReviewGenerativeModels,ReviewGAN}  
aim at modelling complicated probability distributions of data in a way that they can readily be used to generate new samples. 
These techniques model the joint distribution of data, such as images of handwritten digits, and some useful quantities associated with the data, e.g., which of the ten digits is shown.  
The model is then used to generate unseen data by sampling from the learnt joint probability distribution, e.g., produce unseen images of digits. 

In physics, we often start from a Hamiltonian, an action, or just a classical configuration energy, describing the system of interest, and, as such, formally, know the distribution of the elementary degrees of freedom, such as the fields in a field theory or the spin configurations in a classical spin model. Typically, one is interested in studying the behavior of these distributions as a function of tuning parameters, e.g., temperature or coupling constants, and one can think of them as the distribution of data conditioned on these tuning parameters. Since, however, this data is usually very high-dimensional, the essential physical properties can only be captured by evaluating physical quantities, such as symmetry-breaking order parameters and their susceptibilities, or non-local probes of topological properties. In most interesting cases, their evaluation cannot be performed analytically and, hence, numerical techniques have to be used. Among those, in particular, Monte Carlo methods, where observables are estimated by sampling from the data, are powerful, as they, at least in principle, guarantee asymptotic convergence to the true distribution.

Markov chain Monte Carlo (MCMC) techniques work by constructing a first order Markov sequence where the next sample is dependent on the current sample. Unfortunately, these methods can suffer from the problem of large thermalization times and large auto-correlation times (especially near phase transitions), both of which increase drastically with the increase in lattice size.  
For quickly generating uncorrelated samples, we need the auto-correlation time to be small. Starting from a random configuration, for quickly reaching the state of generating valid samples that conform to the underlying true distribution, we need a small thermalization time. Furthermore, MCMC sampling approaches can in practice get stuck in local minima, in spite of being ergodic in theory.

To curtail the effect of dramatic increase of auto-correlation time near criticality, many global update methods have been developed, which simultaneously change the variables at many sites in a single MC update, such as Swendsen-Wang \cite{Swendsen-Wang}, Wolff \cite{Wolf}, worm \cite{worm}, loop \cite{loop1,loop2} and directed loop \cite{directed-loop1,directed-loop2} algorithms. But these methods work only for specific types of models and not for any generic system.

Besides several other promising applications of machine-learning methods in physics \cite{Dunjko_2018,PhysicsToday,MEHTA20191,RevModPhys.91.045002,CarleoRBMReview}, generative modelling techniques have been explored for enhanced generalizability and performance. For instance, Efthymiou and Melko \cite{melkosuperresolution} use deep-learning-based super-resolution techniques to produce spin configurations of larger sizes from MCMC-generated configurations of smaller sizes by the use of convolutional neural networks (CNNs). The resolved configurations have  thermodynamic observables that agree with Monte-Carlo calculations for one and two-dimensional (2D) Ising models. 
Another approach is `self-learning Monte Carlo' \cite{SLMC,SLMC2,SLMC3,RKKYSLMC} 
that, in principle, works for any generic system 
and applies machine learning-based approaches on top of MCMC to speed up the simulations and to reduce the increase in auto-correlation time near the critical temperature. Other approaches which apply machine-learning techniques as a supplement or alternative to MCMC are based on normalizing flow \cite{FlowbasedGenModel}, Boltzmann machines \cite{MelkoTorlai,2017arXiv170804622M,Carleo602,PhysRevB.95.035105}, on reinforcement learning \cite{YingJerIceGeneration}, on generative adversarial networks (GANs) \cite{ScalarFieldTheoryGAN,MillsGANIsing,millsadversarial,PhysRevD.100.011501,IsingDeepGAN,FermiHubbardGAN}, autoencoders \cite{cristoforetti2017towards,1dXYmodel,VAEforPhy}, and on variational autoregressive networks \cite{2019PhRvL.122h0602W,PhysRevLett.124.020503,DingFreeEnergy,IsingGenerative}. 

So far, in most of these approaches, the underlying generative model is trained separately for different values of the tuning parameters of the system, such as different temperatures. But when configurations for multiple temperatures, including close to criticality, need to be generated, either they require configurations for that corresponding temperature and training a model again and/or the Markov chain has to be re-started altogether. For this reason, we here explore a different and less used \cite{PhysRevD.100.011501,IsingDeepGAN,FermiHubbardGAN} strategy, which consists of learning the conditional probability distribution of physical samples, conditioned on tuning parameters.
We train deep-learning-based generative models, including conditional GANs \cite{ConditionalGANs}, over various temperatures which are far from the critical region. Later, we use these models to produce new configurations by providing temperature as input. The model is demonstrated to accurately interpolate across the complete range of temperatures, including the temperatures very close to criticality, over which no training data was provided (interpolation trick). 
The success of such an approach could also allow extrapolation to inaccessible regions of phase diagrams, where no training samples are available since MCMC sampling becomes expensive.  
In addition, we believe that the optimization strategies for generative modeling of physical systems we discuss in this work will also be useful for the application to experimentally generated data \cite{PhysRevLett.123.230504,FermiHubbardGAN}.

Generative models can be broadly subsumed into two categories---prescribed and implicit \cite{ImplicitGAN}. \textit{Prescribed models} are those that provide an explicit parametric specification of the distribution of the output (data). These models typically deploy Bernoulli or Gaussian outputs, depending on the type of data. On the other hand, \textit{implicit models} directly generate data by passing a noise vector through a deterministic function which is generally a neural network. Implicit models can be more expressive than their prescribed counterparts but calculating likelihood becomes intractable in most cases. Most of the generative models in machine learning are prescribed models as they have a notion of likelihood, are easy to optimize and produce excellent results. But, generally, they make an assumption of independence between the parametric distribution across various pixels or lattice sites. Such assumptions in physics can be quite restrictive as the models need to capture the correlations between lattice sites. 
Prescribed models would otherwise need to estimate large co-variance matrices and ensure their positive-definiteness. 
For this reason, we expect and also confirm by our numerical experiments that implicit generative models, in particular in the GAN framework, are more suitable for modelling the site-to-site correlations in physical systems.

Additionally, we propose other modifications that exploit the underlying structure of the physical systems and enhance the model's utility. 
The proposed modifications can bring significant improvement in performance as compared to the prescribed models treated as baselines. We also show that, for implicit models, maximizing the mutual information between a set of structured latent variables and reconstructed configurations leads to maximizing a lower bound on the entropy of the learnt distribution; this reduces the correlations among configurations generated by the model and can act as an indicator of phase transitions.  
We evaluate in detail the improvements in performance of the various modifications we propose. While our approaches can be readily applied to other systems as well, we focus for concreteness in our numerical studies on the 2D XY model, as it provides a transparent example to benchmark these modifications and has been established as a challenging model for neural networks \cite{XYModelSupervised}.

If the type of phase transition and the associated observable, e.g., a local order parameter, are known, these quantities can be evaluated with the generated samples to capture the phase transition. For instance, in case of the XY model, the finite-temperature BKT transition is associated with the proliferation/suppression of vortices \cite{BKTTransition,berezinskii,berezinskiiII,KosterlitzII}. While we show that our generative models can indeed reproduce the expected behavior of vortices, we also demonstrate that our trained network can be used to reveal the transition without requiring knowledge about the underlying nature of the phase transition. This unsupervised detection of phase transitions is another central topic of machine learning in physics. In particular, topological transitions, such as the BKT transition, are challenging due to their non-local nature; however, the method proposed in \refcite{NatPhysTop} has been demonstrated to work in a variety of different models \cite{NatPhysTop,PhysRevLett.124.185501,2020arXiv200202363C} and extensions \cite{PRLTopology} for symmetry-protected topological phases have been developed. We here demonstrate that trained generative models can also be used to indicate the phase transition in an unsupervised way: as expected \cite{vectordiv,InWeightsOfNetworks,WeightOfNetwork,topologicalorder}, we find that the model is particular susceptible to parameter changes in the vicinity of the transition. We quantify this by introducing a fidelity measure constructed on the trained GAN that can be efficiently evaluated and shows peaks in the vicinity of the phase transition.

The remainder of this paper is organized as follows. In \secref{sec:level2}, we provide an introduction to the  different generative modelling techniques we explore in this work and to the XY model. The modifications we propose for an effective modelling of physical systems are described in detail in \secref{sec:level4}. The numerical experiments, using the XY model as concrete example, are presented in \secref{sec:level5}. Finally, \secref{sec:level6} contains a brief summary.

\section{ Generative modelling and XY model}\label{sec:level2}
To establish notation and nomenclature, we first provide an introduction to the generative machine-learning methods we use---variational autoencoders (VAEs) and GANs, as well as their conditional extensions; we also define the 2D XY model, which is the model we use to benchmark our machine learning approach with, and the physical quantities we study. Readers familiar with the XY model and these generative machine-learning techniques, can skip this section and proceed directly with \secref{sec:level4}. 

\subsection{Variational Autoencoders}\label{VariationalAutoencoderIntro}
VAEs are powerful continuous latent variable models used for generative modelling of a high-dimensional distribution over a given data set, allowing one to sample directly from the data distribution 
\cite{VAEtutorial}. They have shown promising results in producing unseen fake images and audio files which are almost indistinguishable from real data. In its standard form, a VAE consists of an encoder and a decoder. The encoder maps from data space $\boldsymbol{X}$ to a latent space  $\boldsymbol{z} \subseteq \mathbb{R}^D$ and consists of a family of distributions $\mathbb{Q}_{\phi}$ on $\boldsymbol{z}$ parameterized by $\phi$; it is typically modeled by deep neural networks. The decoder consists of a family of distributions $\mathbb{P}_{\theta}$ on $\boldsymbol{X}$ parameterized by $\theta$. As the name implies, the encoder encodes the semantic information present in the data into the latent space. The decoder uses the encoded information in latent space to reconstruct the data. The overall objective is to maximize the likelihood of the data, independently and identically distributed as $P(x) = \int P_{\theta}(x|z)P(z) dz$, where, $x\in \boldsymbol{X}$, $z\in\boldsymbol{z}$, $ P_{\theta}(x|z) \in \mathbb{P}_{\theta}$, and $P(z)$ is the prior distribution, often taken as Gaussian. The likelihood is generally intractable to compute but can be maximized by maximizing the evidence lower bound (ELBO). The ELBO for marginal log-likelihood $P_\theta(x)$ for a data-point $x$ is expressed as
\begin{eqnarray*}
  \log P_\theta(x) \geq  \mathbb{E}_{z \sim Q_{\phi}(z|x)}[ \log  &&P_{\theta}(x|z)]  \\&& - D_{\text{KL}}[Q_{\phi}(z|x)||P(z)],
  \end{eqnarray*}
where $ Q_{\phi}(z|x) \in \mathbb{Q}_{\phi}$. 
The ELBO consists of 2 terms: $(i)$ a loss term accounting for the error in the reconstructed data and $(ii)$ a regularizing term which makes the encoder to encode information such that its distribution is close in Kullback-Leibler (KL) divergence, $D_{\text{KL}}$, to the prior distribution. 

\begin{figure}
    \subfloat[Conditional VAE. \label{fig:awesome_imagecvae}]{%
        \includegraphics[width=0.45\textwidth]{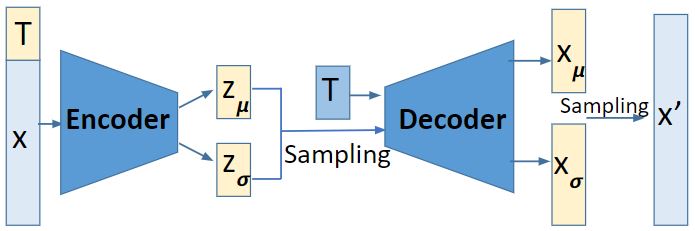}}
    \hfill
    \subfloat[Conditional GAN.\label{fig:awesome_imagecgan}]{%
        \includegraphics[width=0.45\textwidth]{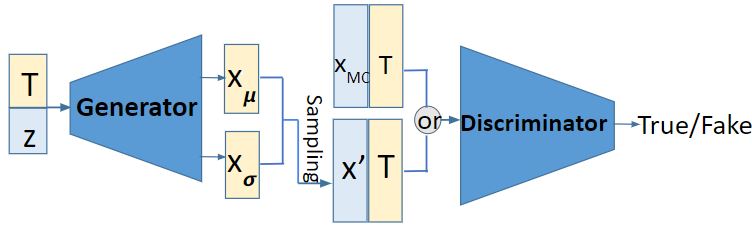}}
    \hfill
    \subfloat[Implicit-GAN. \label{fig:implicitGAN}]{%
        \includegraphics[width=0.45\textwidth]{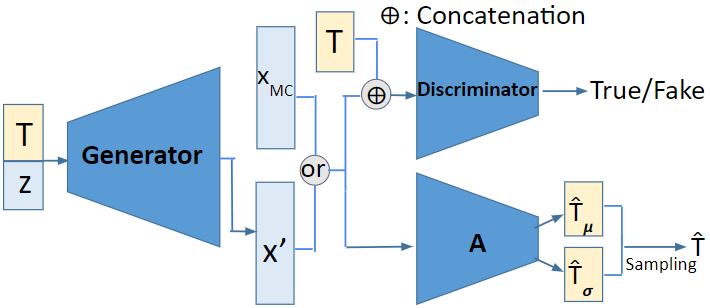}}
    \caption{Block-diagram representation of  (a) C-VAE, (b) C-GAN, and (c), our proposed method, an Implicit-GAN. We refer to the respective parts of the main text, Sec.~\ref{VariationalAutoencoderIntro}, Sec.~\ref{GenerativeAdvNetwIntro}, and Sec.~\ref{sec:level4c}, for a detailed description.}
    \label{fig:awesome_image3}
\end{figure}

\textbf{Conditional VAE (C-VAE)} is a simple extension of standard VAE, with the only difference  that the data distribution as well as the latent distribution are both conditioned by some external information. 
We illustrate the typical structure of a C-VAE in Fig.~\ref{fig:awesome_imagecvae}. The objective is now to maximize the likelihood conditioned on the given information. For our purposes here of generating samples of a physical model, the ``given information'' refers to the tuning parameters of interest in that model, such as temperature, $T$, or ratios of exchange interactions in spin models and so on. Since we will later in Sec.~\ref{sec:level5} focus on temperature, we will use $T$ to denote the given information in the following, but reiterate that it should, in general, be thought of as a multi-component vector comprising several physical tuning parameters.

To train the C-VAE, we again maximize the ELBO, now assuming the form
\begin{eqnarray*}
  \log P_\theta(x|T) \geq  \mathbb{E}_{z \sim Q_{\phi}(z|x,T)}&&[ \log P_{\theta}(x|z,T)] \\&&- D_{\text{KL}}[Q_{\phi}(z|x,T)||P(z|T)].
\end{eqnarray*}
Here, we will assume the prior distribution to be independent of $T $, i.e. $P(z|T) = P(z) = \mathcal{N}(0,I)$. \par

\subsection{Generative Adversarial Networks}\label{GenerativeAdvNetwIntro}
GANs are another powerful framework for modelling a probability distribution. In physics, GANs have been successfully applied to many different models ranging from binary spin systems like the Ising model \cite{MillsGANIsing}, to the Fermi-Hubbard model \cite{FermiHubbardGAN}, high-energy physics \cite{ScalarFieldTheoryGAN}, cosmology \cite{CosmoGAN}, and material science \cite{millsadversarial}. A GAN consists of two models, a generator $G(z)$ and a discriminator $D(x)$. The generator is a function $G:\boldsymbol{z}\rightarrow \boldsymbol{X}$ 
which tries to capture the data distribution and produces samples $x$ that closely resemble samples from the training data. On the other hand, the discriminator is a function $D:\boldsymbol{X}\rightarrow (0,1)$ which tries to estimate the probability that a sample came from the true data distribution 
(true sample) rather than from the generative model $G$ (fake/negative sample). $G$ tries to maximize the probability of $D$ making a mistake while $D$ tries to minimize the probability of being fooled by $G$. The result is a minimax game between two players, described by a value function $V(G,D)$. The objective of this game can be expressed as 
\begin{eqnarray*}
     \min_G \max_D V(G,D) = \min_G \max_D && \  \mathbb{E}_{x \sim p_{\text{Data}}}[\log D(x)] \\&& +  \mathbb{E}_{z \sim p(z) }[\log (1 - D(G(z))].
\end{eqnarray*}

\textbf{Conditional GANs (C-GANs)} are a simple extension \cite{ConditionalGANs} of standard GANs in which the generator produces samples based on the external information $T$ while the discriminator tries to estimate the probability that the sample came from the true conditional data distribution rather than from $G$. The associated minimax objective now becomes
\begin{align}\begin{split}
    \min_G \max_D V(G,D;T) =  \min_G  \max_D \  \mathbb{E}_{x \sim p_{\text{Data}}}[\log D(x;T)] \\ + \mathbb{E}_{z \sim p(z) }[\log (1 -  D(G(z;T);T)]]
\label{OriginalTraningObj}\end{split}\end{align}
and we show the basic structure of a C-GAN in Fig.~\autoref{fig:awesome_imagecgan}.

\subsection{2D XY model}
While the methods we propose and compare in this work are more generally applicable, we will employ one specific physical model, the classical 2D XY-spin model, to illustrate and test the generative machine-learning methods. The XY model was chosen as it features key challenges---compact local degrees of freedom (two-component units vectors) and non-local, topological excitations (vortices) together with conventional excitations (spin waves)---in a minimal setting.  

More specifically, the XY model consists two-component spins on every site $i$ of the lattice with fixed magnitude, which we set to $1$ and, hence, are described by the unit vectors $\vec{s}_i = (\cos\theta_i,\sin\theta_i)^T$, $\theta_i \in [0,2\pi)$. We here consider a 2D square-lattice of size $N \times N$ and restrict ourselves to ferromagnetic nearest-neighbor interactions, $J>0$; using the latter as unit of energy, $J\equiv 1$, the energy of a configuration $\boldsymbol{\theta}=\{\theta_i\}$ is given by
\begin{equation}
    E(\boldsymbol{\theta}) = -\sum_{\langle i,j \rangle}\vec{s}_i \cdot\vec{s}_j = - \sum_{\langle i,j \rangle}\cos(\theta_i - \theta_j), \label{ConfigurationEnergy}
\end{equation}
where the sum over $\langle i,j \rangle$ includes all the adjacent sites on the lattice.

The probability density of a configuration $\boldsymbol{\theta}$ at a given temperature $T \in \mathbb{R}^+$ is given by 
 \begin{equation}
     P_T(\boldsymbol{\theta}) = \frac{1}{Z(T)} e^{-\frac{E(\boldsymbol{\theta})}{ T}}, \label{ProbabilityDistributionOfSamples}
 \end{equation}
where the Boltzmann constant is set to unity and $Z(T) = \sum_{\boldsymbol{\theta}} e^{-\frac{E(\boldsymbol{\theta})}{T}}$ is the partition function. Thermal expectation values, $\braket{\mathcal{O}}_T$, of physical quantities $\mathcal{O}=\mathcal{O}(\vec{\theta})$, such as mean magnetization, $\vec{m}(\vec{\theta}) = N^{-2}\sum_i \vec{s}_i(\theta_i)$, or mean energy, $N^{-2}E(\vec{\theta})$, follow from \equref{ProbabilityDistributionOfSamples} as
\begin{eqnarray}
   \braket{\mathcal{O}}_T = \sum_{\vec{\theta}} \mathcal{O}(\vec{\theta}) P_T(\boldsymbol{\theta}). \label{ExpectationValueOfObs}
\end{eqnarray}
In general, \equref{ExpectationValueOfObs} cannot be evaluated exactly and, hence, has to be analyzed with approximate analytical techniques or numerical approaches.
One of the most common ways of evaluating the sum in \equref{ExpectationValueOfObs} numerically, proceed via MCMC sampling of configurations $\vec{\theta}$ according to the distribution $P_T(\boldsymbol{\theta})$, e.g., via the Metropolis-Hastings algorithm \cite{MH_algo}. 

The goal of this work is to investigate how generative models can be used to generate samples $\vec{\theta}$ for efficient evaluation of the expectation values of observables in \equref{ExpectationValueOfObs}. Besides the mean energy and magnetization mentioned above, we also investigate the number of vortices in the system at a given temperature. Vortices are non-local excitations defined by a non-zero winding, $\nu\neq 0$, of the unit vector $\vec{s}_i$ on any closed path encircling the core of the vortex. Proliferation or suppression of vortices are the defining feature for the finite-temperature phase transition, the BKT transition \cite{BKTTransition,berezinskii,berezinskiiII,KosterlitzII}, of the 2D XY model (\ref{ConfigurationEnergy}). Studying vortices is not only motivated by the fact that they are integral to the physics of the XY model, but also due to their non-local, topological nature; as a consequence, one might expect that vortices are more difficult to capture by machine-learning techniques than local excitations.

In practice, we detect vortices in samples by counting, for every site $i$, the angle differences in anti-clockwise sense around the $(3\times3)$ square centered at $i$. Each difference was constrainted to lie in $(-\pi,\pi]$ using a saw function. The ``vorticity $V$'' of a configuration $\vec{\theta}$ is the number of vortices with winding number $\nu=+1$.

\section{\label{sec:level4}Proposed method}
Having introduced the basic generative models, we will next discuss our proposed implementation and modifications for improved performance on generating samples for the evaluation of physical observables.  
These modifications are motivated from the structure of the physical system. To be concrete, we will discuss them mostly in the context of the 2D XY model, although they apply equally well to many other systems as well. To analyze how relevant the different modifications are, we will perform an ablation analysis in Sec.~\ref{sec:ablation}.

\subsection{Representation of physical states}
The first set of modifications concerns the representation of states. As we will see, choosing a proper way of parameterizing the physical states is integral to an efficient and feasible generative modelling. 

\subsubsection{Exploiting symmetries}\label{sec:level4a}
First of all, many physical systems exhibit symmetries. Formally, this means that the energy $E(x)$ of any state $x$ is the same as that of the transformed state, $x'$, $E(x)=E(x')$. This can be exploited to find a more compact representation of the state: one can represent states such that two states that are related by a symmetry have the exact same representation. Unbiased sampling is guaranteed by randomly performing symmetry transformations on the generated state, since $E(x)=E(x')$ implies that any two symmetry-related states are equally likely.

In the case of the XY model, the symmetry is global rotation of all spins, 
\begin{subequations}\begin{equation}
    \vec{\theta} \quad \longrightarrow \quad \vec{\theta}'= \vec{\theta} + \theta_0, \quad \theta_0\in\mathbb{R}. \label{SpinRotationSymmetry}
\end{equation}
This symmetry allows us to reduce the dimensionality of the representation of the states from $N^2$ to $N^2-1$.
In practice, for any given state $\vec{\theta}$ we choose $\theta_0$ such that 
\begin{equation}
    \left(\vec{m}(\vec{\theta}'_i)\right)_y = N^{-2}\sum_i \sin (\theta'_i) = 0,
\end{equation}\label{AdjustingTheGlobalOrientation}\end{subequations}
i.e., describe the state by deviations of the spin orientations about a certain `mean-direction' (here chosen along the $x$-axis).
As $E(\vec{\theta})$ for the XY model (\ref{ConfigurationEnergy}) is invariant under \equref{SpinRotationSymmetry}, we know $P_T(\boldsymbol{\theta})=P_T(\boldsymbol{\theta}', \theta_0)=P(\boldsymbol{\theta}')P(\theta_0)$, with uniform $P(\theta_0)$. We will model $P(\boldsymbol{\theta}')$ using a deep generative model, and sample $\theta_0$ uniformly in $[0,2\pi)$. Thus, we have reduced the dimensionality of space (the degrees of freedom of data) in which the manifold of lattice configurations is embedded and made the learning simpler.

\subsubsection{Topology of degrees of freedom\label{sec:level4b}}
For many physical systems, the degrees of freedom on every site are compact. For instance, for XY-spin or Heisenberg-spin models, the local configuration space is a one-dimensional or two-dimensional sphere, respectively. In these cases, one has to be careful about choosing a smooth representation of these spaces that respects their topology.

For the XY, the angles $\theta_i \in [0,2\pi)$ have discontinuous jumps at $2\pi$. As such, directly using angles as input to the model does not explicitly take into account the topological and geometrical properties of the space of XY spins. For example, an angle of 2$^{\circ}$ is quite similar to 358$^{\circ}$, and also 180$^{\circ}$ is not a good estimate of the mean spin orientation. The topology at each lattice site can be taken into account by using a two-channel lattice consisting of cosines and sines of lattice angles at both input and output of our model; this means that instead of $\theta_i$, we use the two-component unit vectors $\vec{s}_i = (\cos\theta_i,\sin\theta_i)^T$, as has previously been implemented for different machine learning studies of the XY model (see, e.g., \refcite{XYPCA}).

Such a choice of input and output makes the model 
an implicit model. This also allows to overcome the limitations on the model's ability to capture correlations between lattice sites due to independent sampling from $N(\mu_i,\sigma_i)$ at each lattice site $i$. We use this representation for GAN framework. A similar extension to VAE framework makes the ELBO intractable. Although there exist approaches like that of Ref.~\onlinecite{ImplicitVAE} to overcome this issue, but most of them are based on adversarial training (or likelihood free inference).
\subsubsection{Periodic boundary conditions}
As we are interested in the bulk properties of the XY model and not in the behavior around edges, we will assume periodic boundary conditions throughout this work. Mathematically, this means that we replace $\theta_{(i_1,i_2)}$, by $\tilde{\theta}_{(i_1,i_2)}:=\theta_{((i_1)_N,(i_2)_N)}$, where $(i)_N$ denotes $i$ modulo $N$.
For the implementation with deep neural networks, we increase the size of the lattice from $N\times N$ to $(N+2)\times(N+2)$, keeping the middle $N\times N$ lattice sites the same and filling the sites at the new edges in accordance with the periodic boundary conditions. We expect that this improves the performance of feature extracting kernels of CNN especially at the edges of a lattice.
We use periodic padding of size 1 on the input layer of the encoder (for VAE) or discriminator (for GAN).

\subsection{Proposed conditional  Models\label{sec:level4c}}
Since our main goal is to produce unseen configurations for any $T$, we use conditional VAEs and conditional GANs.
During training, the input to the model (encoder of VAE or discriminator of GAN) at a time is a spin configuration, represented either by a set of angles $x = \{\theta_i\}\in \mathbb{R}^{N^2}$ or two-component unit vectors $x = \{\vec{s}_i\}\in \mathbb{R}^{2N^2}$, and the temperature $T\in\mathbb{R}^+$, which it was sampled at using MCMC. For the ease of implementation with standard CNN libraries, the input is formatted as two channels, one consisting of the configuration $x$ and the other consisting of $T$. This format has also been used by AlphaGo \cite{Alphago}.
After training the model, during testing, we sample $z\sim \mathcal{N}(0,I)$ and feed it into the decoder (in case of the VAE) or the generator (in case of the GAN), along with $T$. For brevity, we will refer to the VAE decoder also as a generator in what follows, and denote the output as $G(z,T)$. 

To further improve the performance of the model, we will make the following two additional modifications, leading to the model which we will call ``ImplicitGAN''.

\subsubsection{Minimizing output biases\label{sec:level4d}}
As mentioned above, we propose to normalize the spin configurations such that their net magnetization vector $\vec{m}(\vec{\theta}')$ always points along the x-axis, see \equref{AdjustingTheGlobalOrientation}. 
But, there is nothing in the training objective (\ref{OriginalTraningObj}) which explicitly incentivizes the network to produce configurations with their magnetization to point along the x-axis. 
If this condition is not satisfied, it implies that our model has developed some bias, which may be due to the model parameters being stuck in a local minimum during training. We indeed observed that the training objective in \equref{OriginalTraningObj} can lead to bad local optima, as discussed later in Sec.~\ref{sec:ablation}. Thus, if we add a term forcing the generative model to minimize the square of the $y$-component of the magnetization in a configuration we can minimize such biases. The GAN training objective now becomes
\begin{eqnarray}
V_b(G,D,T)&& =V(G,D,T)\nonumber\\
        &&+\lambda\min_{G}  \mathbb{E}_{z \sim p(z); \boldsymbol{\theta}'=G(z|T)}\left[\sum_i \frac{\sin(\theta'_i)}{N^2}\right]^2
        \label{eq:ganbias}
\end{eqnarray}
where, $\lambda\in \mathbb{R}^+$ is a constant hyper-parameter.

\subsubsection{Maximizing the output entropy}\label{sec:level4eE}
Generative models only learn to approximate the distribution. Thus, the generated samples will hardly have any practical significance if we cannot guarantee convergence to the exact distribution---especially considering the fact that GANs are susceptible to the mode-collapse problem, i.e., they might miss a subset of the modes of a multimodal distribution of the samples. Still, in practice, we could use the generated $x$ as the initial configuration for MCMC. 
But if the different samples generated by our model have high correlations among themselves, the number of MCMC steps needed to obtain uncorrelated samples would be large, thereby, defeating the purpose of the extra computational efforts for training the generator. We can decrease the number of MCMC steps needed if we can reduce the initial correlation among the different samples generated by our model. 

To achieve this, we propose to additionally maximize the overall entropy of the learnt distribution $h(G(z,T))$, i.e., to make the learnt distribution more `diffused', while also keeping the distribution of generated samples in close agreement to the true distribution for all temperatures. It has been shown that, in case of prescribed models, the entropy-regularized loss function reduces the problem of mode-collapse \cite{PresGAN}. 
In practice, the problem is that $h(x)$ is difficult to compute or maximize. However, we can instead maximize a lower bound on $h(x)$ in the following way: due to the symmetry, $I(x;T) = I(T;x)$, of the mutual information $I$, it holds
\begin{eqnarray*}
    h(x)&& = h(T) - h(T|x) + h(x|T) \\
        && \geq h(T) -  h(T|x) + h(x|T,z). 
\end{eqnarray*}
Now, $h(x|T,z)=0$ for an implicit model (as opposed to prescribed models), because the value of $x$ is completely determined by the value of $\{T,z\}$.  Thus,
\begin{eqnarray}
   h(x) \geq h(T) - h(T|x) = I(T;x).
\end{eqnarray}
Here $h(T)$ is constant because we have already specified and fixed the latent distribution $P(T)$, which is a uniform probability mass function over temperatures in the training data. Hence, minimizing $h(T|x)$ maximizes the lower bound on $h(x)$. Minimizing $h(T|x)$ requires access to the posterior $P(T|x)$. But, we can minimize an upper bound on $h(T|x)$ by defining an auxiliary distribution $A(\hat{T}|x)$ as:
\begin{eqnarray}
    h&&(T|x) = - \mathbb{E}_{x}[\mathbb{E}_{T\sim P(T|x)}[\log P(T|x)]] \nonumber\\
    && = -\mathbb{E}_{x}[D_{\text{KL}}(P(\hat{T}|x)||A(\hat{T}|x)) + \mathbb{E}_{\hat{T}\sim P(T|x)}[\log A(\hat{T}|x)]] \nonumber \\
    && \leq - \mathbb{E}_{x}[\mathbb{E}_{\hat{T} \sim P(T|x)}[\log A(\hat{T}|x)]] \nonumber \\
    && = -\mathbb{E}_{\hat{T}\sim P(T)}[\mathbb{E}_{x\sim P(x|T)}[\log A(\hat{T}|x)]] \nonumber \\
    && \equiv L_H(G,A) \label{UpperBoundONhTx}
\end{eqnarray}
We use an auxiliary network $A$ to estimate the temperature from $x$, i.e., maximize the probability $P(\hat{T}=T)$. Such a technique of maximizing a lower bound on mutual information in terms of an auxiliary distribution was previously proposed in \cite{Infogan}. 
According to \equref{UpperBoundONhTx}, $h(\hat{T}|x)$ can be minimized by minimizing its upper bound given by $L_H(G,A)$. Note the bound becomes tight when $\mathbb{E}_{x}[D_{\text{KL}}(P(\hat{T}|x)||A(\hat{T}|x))] \rightarrow 0$.
The new objective function in terms of this auxiliary distribution becomes:
\begin{eqnarray}
    \min_{G,A} \max_D \{V_b(G,D;T)+ \gamma L_H(G,A)\},
    \label{eq:infogan}
\end{eqnarray}
where $\gamma\in \mathbb{R}^+$ is a constant hyper-parameter. Note, $L_H(G,A)$ maximizes only a lower bound on the entropy and, hence, $h(x)$ is not guaranteed to increase. The gap $h(x|T)-h(x|T,z)=I(x;z|T)$ is expected to be small since, by the structure of model, one does not expect large mutual information between noise variables and generated samples. Since $I(x;z|T)\geq 0$, the overall entropy is likely to increase in practice. Typically, $A$ and $D$ are implemented as neural networks sharing most of the layers. But, in our case, the information of temperature should only be given to $D$ and not $A$, hence they were employed as separate neural networks, as shown in Fig \ref{fig:implicitGAN}. The discriminator $D$ tries to predict the probability that the sample belongs to the true distribution, while the auxiliary network $A$ outputs a distribution over temperatures for a given configuration. The distribution is assumed to be Gaussian with parameters $\hat{T}_\mu$ and $\hat{T}_\sigma$. 

\subsection{\label{sec:level4f}Unsupervised detection of phase transitions}
So far our focus has been to generate samples for the evaluation of physical observables according to \equref{ProbabilityDistributionOfSamples}. If we are interested in studying phase transitions and know which observables capture the transition, e.g., a local order parameter in case of a conventional, symmetry-breaking phase transition, we can simply evaluate these observables with our generated samples. However, one of the central questions of machine learning in the context of condensed matter and statistical physics is to find ways of detecting the transition without ``telling'' the algorithm which observables are relevant. This, in this sense ``unsupervised'', detection of phase transitions could potentially be useful in cases where the order parameter or topological invariant characterizing the transition are not known. 

Having constructed models that can generate samples at a given temperature $T$, we here analyze whether the behavior of these networks as a function of the tuning parameter $T$ can be used to infer where phase transitions take place, without requiring knowledge about which observables to evaluate. In line with previous works \cite{vectordiv,InWeightsOfNetworks,WeightOfNetwork,topologicalorder}, dealing with different machine-learning setups, we expect that our generative models are particularly susceptible to changes in $T$ in the vicinity of phase transitions. 

The first measure we use is directly related to the one defined in previous works \cite{vectordiv,topologicalorder} and makes use of the auxiliary network $A(x)=\hat{T}$ that we implemented to estimate the temperature from the samples $x$, needed to maximize the output entropy. Since one expects $A$ to perform worst in the vicinity of the transition, 
\begin{equation}
    \mathcal{D}(T')= \frac{\partial \hat{T}}{\partial T}\bigg|_{T=T'} \approx \frac{\mathbb{E}[A(x_{T'+\Delta T})]-\mathbb{E}[A(x_{T'-\Delta T})] }{2\Delta T} \label{FirstMeasure}
\end{equation}
should be peaked around the critical temperature.

The second measure we introduce is unique to GANs and can be defined for any GAN architecture, not only for the modified version with the additional auxiliary network. This measure is analogous to the widely studied quantum fidelity, which has also been extended to finite temperature and thermal phase transitions \cite{PhysRevE.79.031101}. It is based on the idea that the form of a state (density matrix for thermal ensembles) will change most dramatically upon modifying a tuning parameter by a small amount (such as temperature $T\rightarrow T+\Delta T$) in the vicinity of a phase transition. This will first require a measure of similarity of two states or ensembles. For this we will use the expectation value of $D(x,T)$ with $x$ taken from some given ensemble $p'$. Since $D(x,T)$ estimates the probability of $x$ coming from the true thermal ensemble, this expectation value quantifies how similar the thermal ensemble and $p'$ are. Since we are interested in shifting the temperature, we replace $p'$ by the ensemble generated by the generator at a different temperature and, thus, define the \textit{GAN fidelity} as
\begin{align}\begin{split}
        \mathcal{F}_{\text{GAN}}(T) = \frac{1}{\Delta T} &\mathbb{E}_{z\sim p(z)}[D(G(z;T),T) \\ &\quad -D(G(z;T),T+\Delta T)].
\label{SecondMeasure}\end{split}\end{align}
Imagine starting in the high-temperature phase and gradually decreasing $T$. Once $T$ reaches the phase transition, the generator in the second term in \equref{SecondMeasure} starts producing samples that are not ``expected'' by the discriminator. Thus, the latter decreases its value, $\mathcal{F}_{\text{GAN}}(T)$ increases, and is expected to peak in the vicinity of the phase transition.
We emphasize that the GAN fidelity in \equref{SecondMeasure} is defined entirely in terms of the networks and can be evaluated very efficiently, once the networks have been trained.

\section{\label{sec:level5} Numerical Experiments}
In this section, we present a detailed study of the performance of the algorithms outlined above using the 2D XY model as a concrete example.
We first compare the proposed method with certain baseline approaches that we define below. 

In the second set of experiments, we train our model over the configurations with temperatures that are below and above the critical temperature. We then test our model over the complete range of temperatures (interpolation trick), i.e., investigate how well it can interpolate over unseen temperatures near criticality. 
In the third set of experiments, we test the ability of our model to detect phase transitions by detecting peaks in the divergence of the model's prediction. 
Finally, we also present an ablation analysis that systematically examines the effectiveness of different components of the proposed method.

\subsection{Generation of training data}
In this work, we use lattices of size $N\times N$, where $N=\{8,16\}$. The training data is obtained using the Metropolis-Hastings algorithm for 32 uniformly spaced values of temperature $T$ in the range $[0.05,2.05]$. For each value of $T$, 10000 configurations
are generated. Starting from a randomly initialized state for each $T$, a sufficiently large number of configurations are rejected initially, to account for thermalization. A configuration is included in the training data set after every 120 MCMC steps for $8\times 8$ and after 400 steps for $16 \times 16$ lattice, to reduce correlations in the training data. The angle at each lattice site is scaled down linearly from $[0,2\pi)$ to $[0,1)$. Thus each configuration is a 2D matrix with each entry between $[0,1)$. The data is then characterized by computing observables like magnetization $\vec{m}$, energy $E$, and vorticity $V$, all as a function of $T$. The samples generated via MCMC as well as the estimated observables serve as the ground truth for evaluations.

\subsection{\label{sec:level5b}Evaluation metrics}
How do we know whether the ensemble of reconstructed configurations statistically belong to true distribution? To evaluate, we compute the aforementioned observables using the reconstructed configurations, and compare the distribution of these observables with the distribution of those generated using MCMC simulations. To compare these distributions, we deploy the following measures on the histograms of observables generated for 500 different configurations.

\subsubsection{Percentage overlap ($\%$OL)}
Our first measure is $\%$OL, which corresponds to the overlap between two histograms, each of which is normalized to unit sum. Mathematically, the $\%$OL of two distributions $P_r$ and $P_\theta$ is calculated as:
    \begin{equation}
      \%\text{OL}(P_r,P_\theta) =   \sum_i \min(P_r(i),P_\theta(i)),
    \end{equation}
where $i$ is the bin index. We use 40 bins in the range [0,1] for the histogram of magnetization and 80 bins in the range [-2,0] for energy. It is not a self-sufficient measure in the sense that the $\%$OL between the histograms can be quite small even though the computed values of observables are sufficiently close to each other. 

\subsubsection{Earth mover distance (EMD)}
The second measure of the distance between two probability distributions we use is EMD with the following interpretation: if the distributions are thought of as two different ways of piling up a certain amount of dirt, the EMD is the minimum cost of turning one pile into the other. Here, the cost is assumed to be the amount of dirt moved times the distance by which it is moved. The EMD $W(P_r,P_\theta)$, between two histograms, $P_r$ and $P_\theta$ of a scalar observable $y$, is defined as 
\begin{eqnarray*}
W(P_r,P_\theta) = \sum_{x=-\infty}^{\infty} \bigg|\sum_{y=-\infty}^x (P_r(y) - P_\theta(y))\bigg|
\end{eqnarray*}

\subsection{Models for Comparison\label{Sec:baselines}}
We perform a series of numerical experiments to test the effectiveness of the proposed methods.  
For comparison, we use modifications of the method of \cite{cristoforetti2017towards} as our two baselines, which provide a reference for the performance of our proposed Implicit-GAN approach.

\subsubsection{C-HG-VAE}
The first baseline model we use is C-HG-VAE. It is a prescribed generative model and was proposed in \cite{cristoforetti2017towards}, referred to by them as HG-VAE, and the only available generative model which tries to improve the reconstruction task for the XY model; as such, it is the most natural starting point for us to construct a baseline model. The C-HG-VAE  employs CNNs instead of fully connected networks to account for translational symmetry of the physical system. To improve the agreement of thermodynamic observables with the ground truth, they modify the standard VAE loss function by additionally including the following $\mathcal{L}_H$ term:
\begin{eqnarray}
   \mathcal{L}_H = (E_{\boldsymbol{\theta}^{\phantom{'}}} - E_{\hat{\boldsymbol{\theta}}})^2,
\end{eqnarray}
where $E_{\boldsymbol{\theta}}$ and $E_{\hat{\boldsymbol{\theta}}}$ are the energies per lattice site of the ground truth and the generated configurations, respectively. Multivariate standard normal distribution was chosen as the prior $P(z)$ and the spin configurations are represented as $\vec{\theta}=\{\theta_i\}$. The output of the decoder (i.e., reconstruction layer) is split into two terms $\mu$ and $\sigma$ corresponding to the parameters of a Gaussian distribution. Configurations were generated by sampling from the Gaussian $\mathcal{N} (\mu_i,\sigma_i)$, $\mu \in\mathbb{R}^{N\times N}, \sigma \in\mathbb{R}^{N\times N}$, with each lattice site $i$ distributed independently. In the abbreviation HG-VAE, $H$ refers to the $\mathcal{L}_H$ term and $G$ to the Gaussian parametric specification of the reconstruction layer. HG-VAE generates new configurations using $z$ sampled from the approximately learned variational distribution $Q_{\phi}(z|x)$ and then feeds these $z$ to the decoder. Generating $z$ from $Q_{\phi}(z|x)$ requires use of MC samples for that corresponding temperature. Hence, their method cannot generate configurations for temperatures not in training data.  But since our goal is to generate configurations even for temperatures for which no training data is available, we modify their method to a conditional model named C-HG-VAE by providing additional information of temperature to both encoder and decoder. For generating new configurations, we provide $z \sim \mathcal{N}(0,I)$ and $T$ to the decoder. $T$ is concatenated multiple times with $z$ so as the decoder does not ignore this information along with multiple $z$. The block diagram representation of C-HG-VAE is the same as that of the C-VAE in Fig.~\ref{fig:awesome_imagecvae}.

\subsubsection{C-GAN}
As second baseline model, we use a prescribed form of a standard C-GAN, introduced in Sec.~\ref{GenerativeAdvNetwIntro}. The C-GAN employing CNNs was trained on the space of angles to reconstruct configurations, given $T$. The input to the generator consists of $T$ concatenated with $\boldsymbol{z} \in \mathbb{R}^N$ sampled from a Gaussian prior, where $N$ is the lattice size. Similar to C-HG-VAE, the generator outputs $\boldsymbol{\mu}_i\in\mathbb{R}^{N\times N}$ and $\boldsymbol{\sigma}_i\in \mathbb{R}^{N\times N}$ corresponding to the parameters of a Gaussian distribution from which the configurations are sampled. The reparametrization trick \cite{VAEtutorial} is used to ensure differentiability of the network. The input of the discriminator has two channels---one consisting of the spin configurations $x$ and the other of $T$. The output of the discriminator is a scalar distinguishing the real from the fake sample.\par

\subsubsection{ImplicitGAN}\label{SummaryOfImplicitGAN}
This is the proposed implicit C-GAN approach. While all of the key components of this method have been motivated and explained in detail in Sec.~\ref{sec:level4c} above, we here provide a concise summary of it:
\begin{enumerate}
    \item The angles $\theta_i$ of the spins in each sample are shifted, $\theta_i \rightarrow \theta_i + \theta_0$, such that the net magnetization vector ($\vec{m}$) always points in the direction corresponding to $\theta_i=0$.
    \item The reconstruction layer of generator consists of two channels $[x_i,y_i]$ with $([x_i,y_i]/\sqrt{x_i^2+y_i^2})$  normalizing function applied at each lattice site. The input of discriminator has 3 channels, with the first two channels consisting of cosines and sines of lattice angles and the 3rd channel containing temperature.
    \item To take into account the periodic boundary conditions of the lattice, we use periodic padding of size $1$ for the input layer of the discriminator.
    \item To minimize the biases, \equref{eq:ganbias} was used as objective function. The value of $\lambda$ was chosen to be $10$ for $8 \times 8$ and $1$ for $16 \times 16$ lattices. 
    \item To maximize the entropy of generated samples, the output layer of the discriminator now has two outputs, $A(\hat{T}|G(z,T))$ and $D(x)$, with learning objective given in \equref{eq:infogan}. The value of $\gamma$ was chosen to be 100 and 10 for $16\times 16$ and $8\times8$ lattices, respectively.
\end{enumerate}

\begin{figure*}
    \subfloat[Mean magnetization $\langle|\vec{m}|\rangle$ for $(8 \times 8)$ lattice \label{subfig-1:dummy}]{%
        \includegraphics[width=0.4\textwidth]{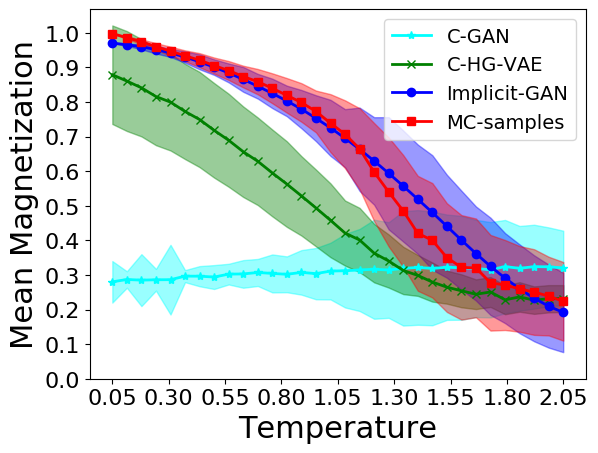}}
    \subfloat[Mean magnetization $\langle|\vec{m}|\rangle$ for $(16 \times 16)$ lattice \label{subfig-13:dummy}]{%
        \includegraphics[width=0.4\textwidth]{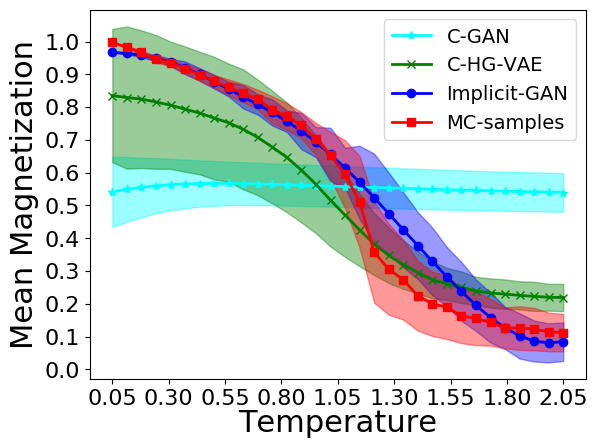}} 
    \\
    \subfloat[Mean energy $\langle E\rangle$ for $(8 \times 8)$ lattice\label{subfig-12:dummy}]{%
        \includegraphics[width=0.4\textwidth]{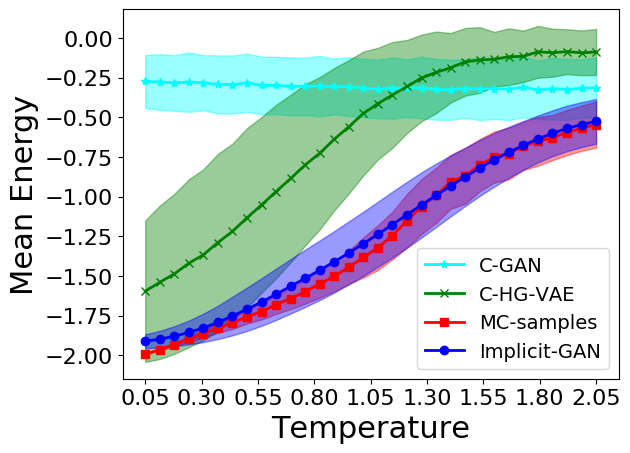}}
    \subfloat[Mean energy $\langle E\rangle$ for $(16 \times 16)$ lattice. \label{subfig-13b:dummy}]{%
        \includegraphics[width=0.4\textwidth]{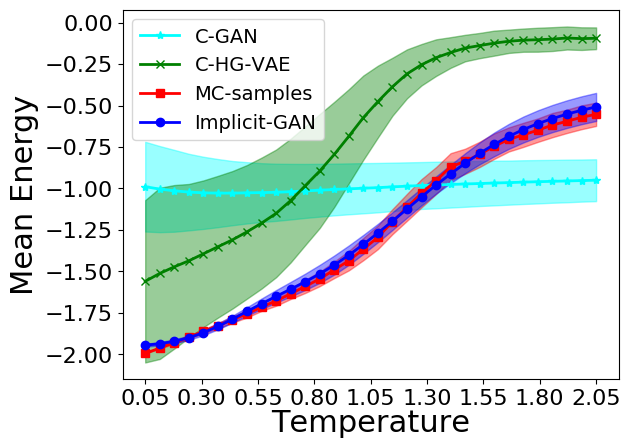}}
    \caption{Mean values of observables computed over lattices generated by different methods, as a function of temperature. Shaded portion indicate standard deviation of the corresponding observable. MC samples are taken as the ground truth; the method giving more overlap with the ground truth is better.}
    \label{fig:awesome_image3}
\end{figure*}
\begin{figure}
    \subfloat[Mean magnetization $\langle|\vec{m}|\rangle$ for $(8 \times 8)$ lattice \label{subfig-41:dummy}]{%
        \includegraphics[width=0.5\linewidth]{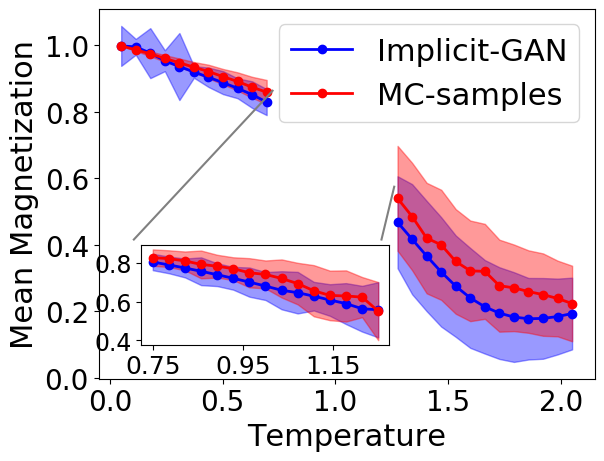}}
    \subfloat[Mean magnetization $\langle|\vec{m}|\rangle$ for $(16 \times 16)$ lattice \label{subfig-42:dummy}]{%
        \includegraphics[width=0.5\linewidth]{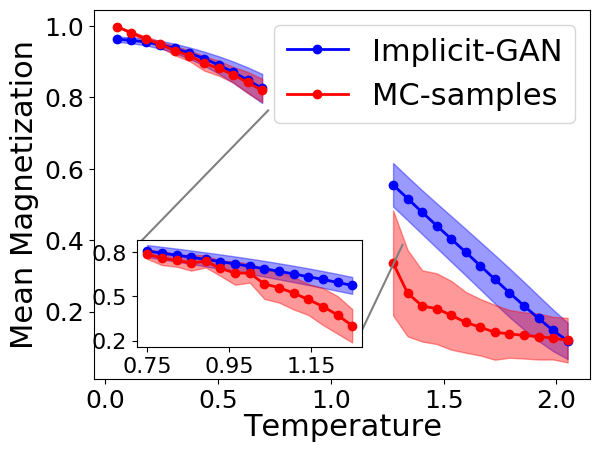}}
    \\
    \subfloat[Mean energy $\langle E\rangle$ for $(8 \times 8)$ lattice\label{subfig-42:dummy}]{%
        \includegraphics[width=0.5\linewidth]{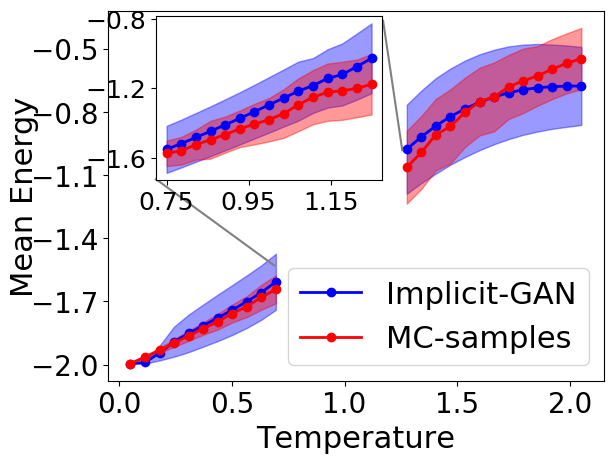}}
    \subfloat[Mean energy $\langle E\rangle$ for $(16 \times 16)$ lattice \label{subfig-42b:dummy}]{%
        \includegraphics[width=0.5\linewidth]{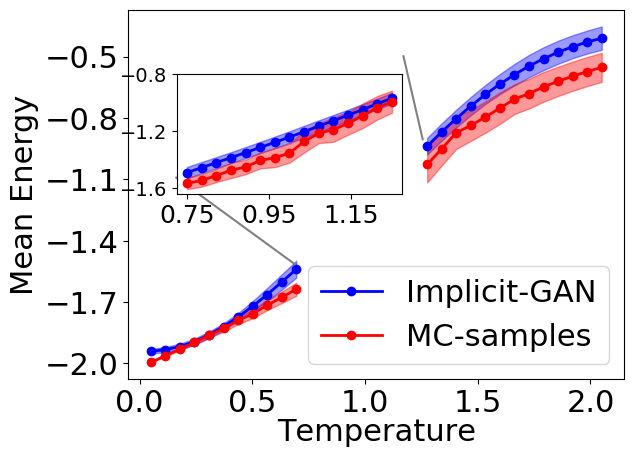}}
    \caption{Mean values of observables computed over samples generated by the proposed method and MCMC (ground truth). The insets correspond to the temperatures which were not part of the training set. These are close to the critical temperature. The shaded portions indicate the standard deviation. More overlap with the ground truth is better.}
    \label{fig:awesome_image5}
\end{figure}
\begin{figure*}
        \subfloat[$\mathcal{D}(T)$ computed across various temperatures. The peaks are observed around the critical temperature. The shaded portion is $0.950\pm0.0625$.
        \label{subfig-phase_trst}]{
        \includegraphics[width=0.33\textwidth]{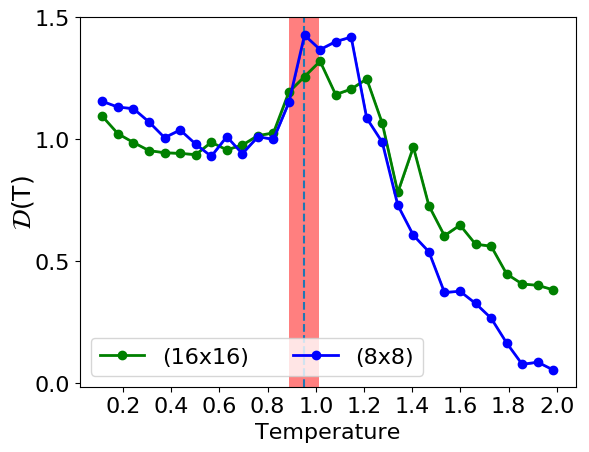}}
    \hfill
    \subfloat[$\mathcal{F}_{GAN}(T)$ computed across various temperatures.\label{subfig-phase_trst2}]{\includegraphics[width=0.33\textwidth]{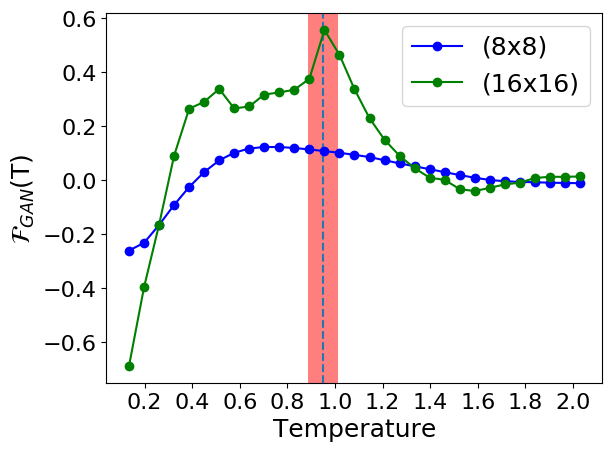}}
    \subfloat[Observed vorticity for $16\times16$ sites as a function of temperature.\label{fig-Vorticity}]{%
        \includegraphics[width=0.33\textwidth]{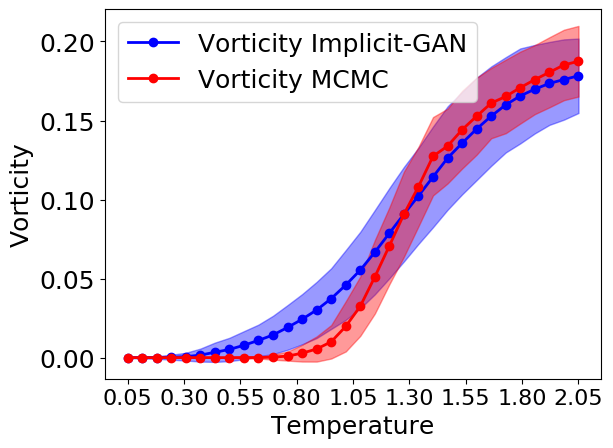}}
    \hfill
    \subfloat[Average value of Y-component of magnetization computed over 500 configurations. \label{subfig-61:dummy}]{%
            \includegraphics[width=0.33\textwidth]{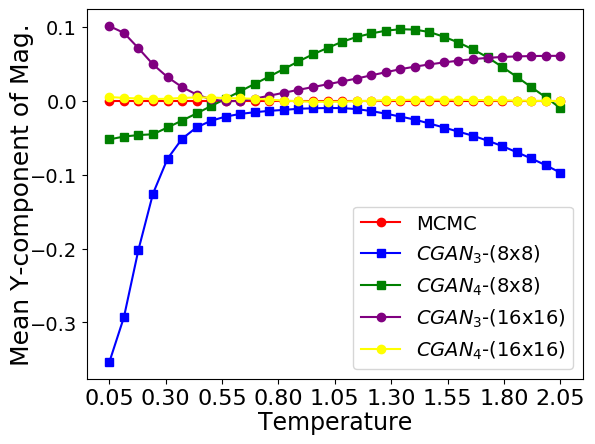}}
    \subfloat[ Average Cross-Correlation between independent configurations generated by model vs.~temperature. MCMC samples generated by Metropolis-Hastings algorithm were used. \label{subfig-cross_corr}]{%
        \includegraphics[width=0.33\textwidth]{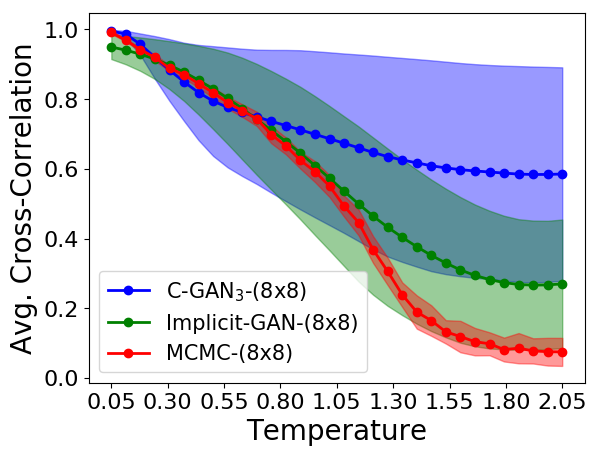}
        \includegraphics[width=0.33\textwidth]{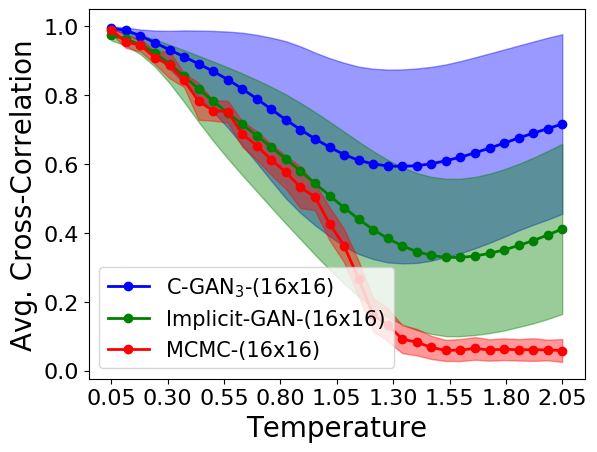}}
    \label{subfig-cross_corr8}
    \caption{}
    \label{fig:awesome_image6c}
\end{figure*}

\begin{table*}
\caption{\label{tab:Table1} Evaluation metrics, as defined in Sec.~\ref{sec:level5b}, along with standard deviation, computed over 500 configurations and averaged across all temperatures. Smaller EMD and higher $\%$OL are better. Best values are indicated in bold.} 
\begin{ruledtabular}
\begin{tabular}{ccccc}
Metric        & Lattice size   & C-GAN        & C-HG-VAE   & Implicit-GAN        \\ \hline
EMD           & $8\times8$     & $ 0.358\pm0.246  $ & $ 0.157\pm0.086  $ & $ \bm{0.038\pm0.024}$  \\
Magnetization & $ 16\times16 $ & $ 0.152\pm0.056  $ & $ 0.118\pm0.028  $ & $ \bm{0.041\pm0.043}$  \\ \hline
EMD           & $ 8\times8   $ & $ 0.484\pm0.250  $ & $ 0.256\pm0.063  $ & $ \bm{0.022\pm0.012}$  \\
Energy        & $ 16\times16 $ & $ 0.233\pm0.140  $ & $ 0.296\pm0.060  $ & $ \bm{0.010\pm0.005}$  \\ \hline
$\%$OL        & $ 8\times8$    & $ 29.31\pm33.35$ & $ 52.18\pm19.15$ & $ \bm{76.69\pm6.46}$ \\
Magnetization & $ 16\times16 $ & $ 7.97\pm16.39$  & $ 42.78\pm17.33$ & $ \bm{67.34\pm20.41}$ \\ \hline
$\%$OL        & $8\times8$     & $ 9.43\pm13.94$  & $ 10.29\pm5.43 $ & $ \bm{68.28\pm20.72}$ \\
Energy        & $ 16\times16 $ & $ 13.64\pm19.33$ & $ 0.62 \pm0.03 $ & $ \bm{73.38\pm19.54}$ 
\end{tabular}
\end{ruledtabular}
\end{table*}
\subsection{Results}\label{ResultsSection}
\subsubsection{Comparison with baselines}

\begin{table*}
\caption{\label{tab:Table2} \textbf{Interpolation Analysis}: Compares evaluation metrics, along with their standard deviation, for C-HG-VAE, Implicit-GAN and Implicit-GAN-Limited. C-HG-VAE, Implicit-GAN were trained over all temperatures ($T\in [0.05, 2.05]$), while Implicit-GAN-Limited was trained over temperatures not in critical zone (critical zone is $T\in[0.75, 1.25]$). Metrics are computed over 500 configurations and averaged across corresponding temperatures. Smaller EMD is better while higher \%OL is better.}
\begin{ruledtabular}
\begin{tabular}{cccccc}
                        & Lattice          & Mean EMD             & Mean EMD             & Mean \%OL           & Mean \%OL           \\
                        & Size             & Magnetization        & Energy               & Magnetization       & Energy              \\ \hline
$T$ outside critical zone &                  &                      &                      &                     &                     \\ \hline
C-HG-VAE                & $(8 \times 8)$   & 0.120 $\pm$ 0.066    & 0.229 $\pm$ 0.047    & 58.217 $\pm$ 18.314 & 10.575 $\pm$ 6.113  \\
                        & $(16 \times 16)$ & 0.108 $\pm$ 0.026    & 0.274 $\pm$ 0.05     & 37.783 $\pm$ 15.703 & 0.0 $\pm$ 0.0           \\
Implicit-GAN            & $(8 \times 8)$   & 0.039 $\pm$ 0.027    & 0.017 $\pm$ 0.011    & 76.058 $\pm$ 6.951  & 67.025 $\pm$ 23.320 \\
                        & $(16 \times 16)$ & 0.041 $\pm$ 0.045    & 0.009 $\pm$ 0.006    & 65.455 $\pm$ 22.817 & 71.666 $\pm$ 21.673 \\
Implicit-GAN            & $(8 \times 8)$   & 0.052 $\pm$ 0.032    & $0.003 \pm 0.001$   & $76.888 \pm 10.610$ & $64.42 \pm 18.512$  \\
limited                 & $(16 \times 16)$ & $0.083 \pm 0.019$    & $0.032 \pm 0.0273$  & $50.041 \pm 27.904$ & $40.333\pm 20.387$  \\ \hline
$T$ inside critical zone  &                  &                      &                      &                     &                   \\ \hline
C-HG-VAE                & $(8 \times 8)$   & 0.268 $\pm$ 0.018    & 0.339 $\pm$ 0.017    & 34.05 $\pm$ 4.728   & 9.425 $\pm$ 2.16    \\
                        & $(16 \times 16)$ & 0.055 $\pm$ 0.018    & 0.365 $\pm$ 0.035    & 57.75 $\pm$ 12.763  & 0.025 $\pm$ 0.066   \\
Implicit-GAN            & $(8 \times 8)$   & 0.033 $\pm$ 0.003    & 0.035 $\pm$ 0.007    & 75.37 $\pm$ 4.211   & 71.071 $\pm$ 8.33  \\
                        & $(16 \times 16)$ & 0.040 $\pm$ 0.038    & 0.012 $\pm$ 0.003    & 73.55 $\pm$ 12.370  & 77.025 $\pm$ 4.670  \\
Implicit-GAN            & $(8 \times 8)$   & $0.0473 \pm 0.0134 $ & $0.038 \pm 0.008$ & $74.559 \pm 5.605 $   & $ 60.66\pm 14.407$  \\
limited                 & $(16 \times 16)$ & $0.0991 \pm 0.0765 $ & $0.031 \pm 0.009$  & $48.426 \pm 16.515$   & $40.986\pm 14.169$  
\end{tabular}
\end{ruledtabular}
\end{table*}

The trained models were tested by computing observables namely magnetization and energy over the reconstructed configurations. Fig.~\ref{fig:awesome_image3} illustrates mean magnetization $\langle|\vec{m}|\rangle$ and mean energy $\langle E\rangle$ values as a function of $T$. We can notice that $\langle|\vec{m}|\rangle$ decreases and $\langle E\rangle$ increases with $T$ for all methods except C-GAN. This shows that C-GAN fails to capture the statistics of the data it is supposed to generate. Also, we can notice that the distribution of Implicit GAN generated observables is closer to the ground truth (MC) as compared to that of C-HG-VAE generated observables. These results, with the metrics averaged across temperatures, are quantified in Table~\ref{tab:Table1}. The implicit-GAN produces the best results over all the metrics as well as lattice sizes. We also note that its performance does not decrease when doubling the linear system size from $8 \times 8$ to $16 \times 16$, which indicates the same method also works for larger system sizes.

\subsubsection{Interpolating unseen temperatures around $T_c$}
After having obtained an architecture capable of modelling the joint distribution of spin configurations across temperatures, we test whether this model can generate samples in the vicinity of the phase transition without having been trained on samples in that regime---a necessary requirement for being able to use generative models to overcome the effect of increase in auto-correlation time in MCMC near criticality.
We define the critical region as $T\in[0.75,1.25]$. Note that the critical temperature is $T_c\approx 0.89$ \cite{CriticalTemperatureBKT} for large system sizes; due logarithmic finite-size corrections, we expect it to be larger, about $0.95$, for our system sizes \cite{XYModelSupervised}.

To test this idea, we train a new Implicit GAN model, named \textit{Implicit-GAN-Limited} with the proposed method on the configurations for temperatures in the interval $[0.05,0.75] \cup [1.25,2.05]$, i.e., outside the critical region. This corresponds to 25\% reduction in training data.  Then we test our model by also interpolating for the temperatures which are not even present in the training data. The metrics over the critical zone were calculated for 15 different temperatures equally spaced in the range $[0.75,1.25]$. The values of the metrics in the critical and non-critical regime are reported in Table~\ref{tab:Table2}. The table shows that there is not much decrease in the accuracy over unseen temperature as compared to temperatures in training data. The model still performs better than the baseline C-HG-VAE, even if the latter was trained on all temperatures, and the performance remains comparable to Implicit GAN trained on all temperatures.  Fig.~\ref{fig:awesome_image5} clearly shows that our model is able to reconstruct configurations even for the temperatures not present in the training data and the observables on these configurations agree well with the MCMC samples. Thus modelling the joint distribution of spin configurations across the temperatures and interpolating it to temperatures not present in the training data is a promising alternative to MCMC sampling for temperatures near the transition, where MCMC techniques can become quite expensive.

\subsubsection{Detecting phase transitions}\label{DetectionPhaseTrans}
We now analyze the ability of the model to detect phase transitions by analyzing its susceptibility to changes in temperature using the two measures introduced in \secref{sec:level4f}.

We begin with $\mathcal{D}$ in \equref{FirstMeasure} which is plotted in Fig.~\ref{subfig-phase_trst} with $\Delta T = 0.0625$,
computed over 500 configurations produced by the generator. We observe that it exhibits peaks in the vicinity of the expected phase transition. However, there is no clear maximum, but rather a double-peak feature. Also the finite-size scaling is opposite to what one would expect, since the double-peak features move to larger rather than smaller temperatures with increasing $N$. More dramatically, the trend does not indicate that these features approach the true location of the transition at large $N$ as they are further way from the BKT transition temperature for larger $N$. A more detailed finite-size scaling analysis would be required to address this issue.

Instead, we here focus on the second measure---the GAN-fidelity---defined in \equref{SecondMeasure} with corresponding plot in Fig.~\ref{subfig-phase_trst2} with $\Delta T = 0.0625$, 
For the larger system size, we here observe a clear, isolated peak very close (around $T\approx 0.95$) to the expected transition temperature for that system size. For the smaller system size, the peak gets broader and is also shifted to the left. While the broadening is a natural feature of smaller $N$, the shift of its maximum is not the expected finite-size scaling trend---this is similar to $\mathcal{D}$, but now seems to approach the correct value with increasing $N$. One reason for the unexpected trend in the peak position could be that $\mathcal{F}_{\text{GAN}}$ is more reliable for the GAN with the larger system size: we found that, at lower $N$, the discriminator is not as successful in determining fake samples (we find $\mathbb{E}[D(G(z;T),T)]$ around $0.45$ for $N=8$ as opposed to around $0.15$ for $N=16$). Note that the negative values of $\mathcal{F}_{\text{GAN}}$ at very low $T$ are clearly unphysical and just related to the fact that the generator underestimates the magnetization slightly at low temperatures, see \figref{fig:awesome_image3}(a,b).

Notwithstanding these issues, it is encouraging to see that we can capture the phase transition without prior knowledge of the underlying relevant observable, using the simple measure $\mathcal{F}_{\text{GAN}}$ that is readily evaluated once the generative model has been trained. Further work, however, is required to see what the advantages and limitations of this approach are and to understand the finite size scaling behavior in the XY and other models. Likely, a combination with unsupervised clustering algorithms, e.g., that of \refcite{NatPhysTop}, can provide additional assistance in detecting phase transitions in an unsupervised way.

On top of being able to capture the phase transition in an unsupervised way, we are dealing with a generative model. Consequently, in cases where we do know the physical quantity capturing the phase transition, we can also directly compute it with the samples generate by the networks. In the case of the 2D XY model, the transition is characterized by the suppression (proliferation) of vortices when entering the low-temperature (high-temperature) phase. For this reason, we have computed the number of vortices as a function of temperature, both in the generated and in the MCMC samples; as can be seen in Fig.~\ref{fig-Vorticity}, we find good agreement. This shows that the Implicit-GAN approach can, indeed, capture topological excitations reliably.

\begin{table*}
\caption{\label{tab:Table3} \textbf{Ablation analysis:} Evaluation metrics, along with standard deviation, computed over 500 configurations of a $16\times 16$ lattice, averaged across all temperatures. Smaller EMD and higher $\%$OL are better. }
\begin{ruledtabular}
\begin{tabular}{lcccccc}
Metric& C-GAN & C-GAN$_1$ & C-GAN$_2$ & C-GAN$_3$ & C-GAN$_4$ & Implicit C-GAN \\ \hline
EMD Magnetization  & 0.304$\pm$0.113   & 0.29$\pm$0.1     & 0.136$\pm$0.04    & 0.071$\pm$0.075  & 0.043$\pm$0.038   & 0.041$\pm$0.043   \\
EMD Energy         & 0.234$\pm$0.14    & 0.212$\pm$0.122  & 0.098$\pm$0.064   & 0.034$\pm$0.028  & 0.041$\pm$0.035   & 0.010$\pm$0.005   \\
\%OL Magnetization & 7.969$\pm$16.394  & 20.6$\pm$21.275  & 41.181$\pm$21.295 & 67.068$\pm$16.092& 69.275$\pm$22.586 & 67.343$\pm$20.415 \\
\%OL Energy        & 16.643$\pm$13.863 & 18.381$\pm$8.303 & 35.269$\pm$23.922 & 47.25$\pm$21.815 & 37.181$\pm$22.397 & 73.382$\pm$19.541 \\

\end{tabular}
\end{ruledtabular}
\end{table*}

\subsection{Ablation analysis\label{sec:ablation}}
We now perform an ablation analysis to examine the effect of each of the components of our proposed Implicit-GAN approach, see Sec.~\ref{sec:level4} and Sec.~\ref{SummaryOfImplicitGAN}, separately. For the sake of comparison, we average the values of the metrics defined in Sec.~\ref{sec:level5b} across all the temperatures used in the training data
and we name our models as 
\begin{enumerate}
    \item C-GAN: The standard prescribed C-GAN, which is also used as a baseline (Sec.~\ref{Sec:baselines}). 
    \item C-GAN$_1$: A standard implicit C-GAN modeling $\theta_i$ using the angles $\theta_i$ rather than the two-component unit vectors $\vec{s}_i$ as input. The generator is a deterministic function of $z$ and outputs the angles $\theta_i$.
    \item C-GAN$_2$: It is same as C-GAN$_1$ model but trained using $\vec{s}_i = (\cos \theta_i,\sin \theta_i)$ as input. It also includes periodic padding of size 1 but the total magnetization of each sample of the training data was not rotated to point along the x-axis. 
    \item C-GAN$_3$: It is same as C-GAN$_{2}$ with magnetization direction normalization as in  \equref{AdjustingTheGlobalOrientation}. 
    \item C-GAN$_4$: same as C-GAN$_{3}$ but the training objective is now modified according to \equref{eq:ganbias}, in order to minimize the output bias. \item Implicit-GAN: This is the proposed implicit C-GAN as was used in Sec.~\ref{SummaryOfImplicitGAN} above. It is the same as C-GAN$_4$ but with the entropy-regularized objective of Sec.~\ref{sec:level4eE}.
\end{enumerate}

The performance of each of these models over the metrics is given in \tableref{tab:Table3}. A comparison between C-GAN and C-GAN$_1$ illustrates that, keeping other factors the same, implicit models perform better than prescribed models. 
Accounting for the continuity of the space of angles and the periodic boundary conditions further improves the performance as can be seen by comparing C-GAN$_1$ with C-GAN$_2$.
Exploiting the global spin-rotation symmetry of the XY model brings further improvement in the agreement of the observables, as is visible from the performance of C-GAN$_3$. 

We see that the performance of C-GAN$_4$ is comparable to C-GAN$_3$ for the metrics in \tableref{tab:Table3}. However, one has to note that these metrics are not directly sensitive to whether the generator satisfies the constraint of total magnetization pointing along the $x$ axis, $\sum_{i}\sin(\theta_i)/N^2=0$; the additional term $\propto \lambda$ in \equref{eq:ganbias} explicitly incentivizes the generator to obey the constraint.
To test this, we compare the average values of the y-component of the magnetization, before (C-GAN$_3$) and after (C-GAN$_4$) adding the term $\propto \lambda$.
Fig.\autoref{subfig-61:dummy} shows a significant reduction in the average `bias', as with C-GAN$_4$ the curves are closer to x-axis. This can be considered as a first-order moment matching test to check whether the model learns the true distribution of the samples, which were reprocessed according to \equref{AdjustingTheGlobalOrientation}.  The parameter $\lambda \approx 1-10$ was observed to work well. With a large value of $\lambda (\approx 100)$, the average bias across temperatures becomes small but the performance of the model over the metrics starts degrading. Hence, there exists a trade-off between the performance and bias. 

Finally, we can see in \tableref{tab:Table3} that the performance of Implicit C-GAN, in terms of reproducing the distribution of observable, is comparable to that of C-GAN$_3$ and C-GAN$_4$ for magnetization and seems to become even better for the energy. On top of that, the key advantage of the Implicit C-GAN is that it generates more uncorrelated samples as compared to the latter.

To quantify this, we measure correlations between a pair of independent samples, $\boldsymbol{\theta}=\{\theta_j\}$ and $\boldsymbol{\theta}'=\{\theta'_j\}$ generated by our models. To this end, we introduce 
\begin{eqnarray}
        \kappa(T) =  \frac{1}{N^2}\sum_j \bigg | \mathbb{E} \bigg[e^{i(\theta_{j} - \theta_0)}e^{-i(\theta'_j - \theta'_0)}\bigg]\bigg| 
    \label{eq:kappa}
\end{eqnarray}
as our measure for the \textit{average cross-correlation}. Here, $\theta_0 = \sum_j (\theta_j/N^2)_{2\pi}$ and $\theta'_0 = \sum_j (\theta'_j/N^2)_{2\pi}$ to make sure that we do not get $\kappa\approx 0$ simply because we have exploited the global spin-rotation symmetry, see \secref{sec:level4a}. The expectation value in \equref{eq:kappa} is taken with respect to the configurations generated by the models and the MCMC. For the latter, we average over all MCMC samples generated with arbitrary separation in the Markov-Chain, i.e., measure the correlation between subsequent samples used to compute observables.
As expected, Fig.~\ref{subfig-cross_corr} shows significant reduction in cross-correlation as compared to C-GAN$_{3}$ for both $8\times8$ and $16\times16$ lattices. Note that the increase of $\kappa$ in all of the curves, including that of the MCMC, at low temperatures  is simply due to the finite magnetization present in finite lattices at sufficiently low temperatures.

\section{\label{sec:level6}Conclusions}
In this work, we have studied different deep-learning based approaches for generating spin configurations. We have discussed in detail several modifications of the basic models in order to have a more efficient representation of the states, that, e.g., takes into account symmetries of the system and the geometry of the local degrees of freedom; the correlations between the samples generated by the model are shown to be reduced by incentivizing our model to increase the entropy of the learnt distribution. A detailed evaluation, via an ablation analysis, of the efficiency of these modifications has been presented. Although the approaches used are more generally applicable, we employed the 2D XY model to benchmark the models' performances. To this end, samples were generated using MCMC to train the models. MCMC was also used to provide the ground truth to compare the generated samples with. For the latter, we investigate the histograms of relevant observables---magnetization, energy, and vorticity. We further quantified the correlations between samples. Overall, we found that implicit models perform better and, in particular, our proposed ImplicitGAN, outperforms all other models considered.  

We have focused on conditional models, which, after training, can be used to generate configurations for arbitrary tuning parameters, in our case temperature. We demonstrate that this can be used to generate configurations near criticality, even without providing training data in the vicinity of the transition. This could be useful for circumventing critical slowing down in MCMC simulations. It also provides the perspective that, instead of storing a huge amount of samples for an interesting model, one could just store a precisely trained neural network to generate samples for future use. We further hope that, when applied to experimental data, it can be used to gain insights about parameter regimes inaccessible in the lab.

Finally, we have also shown that trained networks themselves can be employed to detect phase transitions, without any prior knowledge, by investigating the networks' susceptibility to parameter changes. Most importantly, we propose a GAN fidelity measure that can be readily evaluated for any trained GAN and is demonstrated to peak in the vicinity of transitions, in analogy to the well-known quantum fidelity measure and its thermal extensions \cite{PhysRevE.79.031101}. We hope that this can supplement unsupervised clustering algorithms, such as that of \refcite{NatPhysTop}, for future machine-learning-based studies of phase transitions.

In the future, we are planning to further test and refine the ImplicitGAN model, by applying it to other classical models, and study its potential for quantum mechanical systems.

\begin{acknowledgments}
MS acknowledges support from the National Science Foundation under Grant No.~DMR-1664842.  
\end{acknowledgments}

\vspace{1em}

\textit{Note added}---During the final stages of the completion of this project, another work appeared on arXiv \cite{XYModelVAN}, where a different generative ML technique is applied to the 2D XY model. The emphasis of this work is different from ours and, in particular, does not contain the analysis of implicit and prescribed models, and that of network-based unsupervised indicators ($\mathcal{D}$ and $\mathcal{F}_{\text{GAN}}$) of the phase transition, but instead relies on the helicity modulus.


\begin{thebibliography}{67}%
\makeatletter
\providecommand \@ifxundefined [1]{%
 \@ifx{#1\undefined}
}%
\providecommand \@ifnum [1]{%
 \ifnum #1\expandafter \@firstoftwo
 \else \expandafter \@secondoftwo
 \fi
}%
\providecommand \@ifx [1]{%
 \ifx #1\expandafter \@firstoftwo
 \else \expandafter \@secondoftwo
 \fi
}%
\providecommand \natexlab [1]{#1}%
\providecommand \enquote  [1]{``#1''}%
\providecommand \bibnamefont  [1]{#1}%
\providecommand \bibfnamefont [1]{#1}%
\providecommand \citenamefont [1]{#1}%
\providecommand \href@noop [0]{\@secondoftwo}%
\providecommand \href [0]{\begingroup \@sanitize@url \@href}%
\providecommand \@href[1]{\@@startlink{#1}\@@href}%
\providecommand \@@href[1]{\endgroup#1\@@endlink}%
\providecommand \@sanitize@url [0]{\catcode `\\12\catcode `\$12\catcode
  `\&12\catcode `\#12\catcode `\^12\catcode `\_12\catcode `\%12\relax}%
\providecommand \@@startlink[1]{}%
\providecommand \@@endlink[0]{}%
\providecommand \url  [0]{\begingroup\@sanitize@url \@url }%
\providecommand \@url [1]{\endgroup\@href {#1}{\urlprefix }}%
\providecommand \urlprefix  [0]{URL }%
\providecommand \Eprint [0]{\href }%
\providecommand \doibase [0]{http://dx.doi.org/}%
\providecommand \selectlanguage [0]{\@gobble}%
\providecommand \bibinfo  [0]{\@secondoftwo}%
\providecommand \bibfield  [0]{\@secondoftwo}%
\providecommand \translation [1]{[#1]}%
\providecommand \BibitemOpen [0]{}%
\providecommand \bibitemStop [0]{}%
\providecommand \bibitemNoStop [0]{.\EOS\space}%
\providecommand \EOS [0]{\spacefactor3000\relax}%
\providecommand \BibitemShut  [1]{\csname bibitem#1\endcsname}%
\let\auto@bib@innerbib\@empty
\bibitem [{\citenamefont {Salakhutdinov}(2015)}]{GenerativeModelsReview2015}%
  \BibitemOpen
  \bibfield  {author} {\bibinfo {author} {\bibfnamefont {R.}~\bibnamefont
  {Salakhutdinov}},\ }\bibfield  {title} {\enquote {\bibinfo {title} {Learning
  deep generative models},}\ }\href {\doibase
  10.1146/annurev-statistics-010814-020120} {\bibfield  {journal} {\bibinfo
  {journal} {Annual Review of Statistics and Its Application}\ }\textbf
  {\bibinfo {volume} {2}},\ \bibinfo {pages} {361} (\bibinfo {year}
  {2015})}\BibitemShut {NoStop}%
\bibitem [{\citenamefont {Wang}(2018)}]{GenerativeModelsforPhysicists}%
  \BibitemOpen
  \bibfield  {author} {\bibinfo {author} {\bibfnamefont {L.}~\bibnamefont
  {Wang}},\ }\bibfield  {title} {\enquote {\bibinfo {title} {{Generative Models
  for Physicists}},}\ }\href
  {http://wangleiphy.github.io/lectures/PILtutorial.pdf} {\  (\bibinfo {year}
  {2018})}\BibitemShut {NoStop}%
\bibitem [{\citenamefont {{Ou}}(2018)}]{ReviewGenerativeModels}%
  \BibitemOpen
  \bibfield  {author} {\bibinfo {author} {\bibfnamefont {Z.}~\bibnamefont
  {{Ou}}},\ }\bibfield  {title} {\enquote {\bibinfo {title} {{A Review of
  Learning with Deep Generative Models from Perspective of Graphical
  Modeling}},}\ }\href@noop {} {\bibfield  {journal} {\bibinfo  {journal}
  {arXiv e-prints}\ } (\bibinfo {year} {2018})},\ \Eprint
  {http://arxiv.org/abs/1808.01630} {arXiv:1808.01630 [cs.LG]} \BibitemShut
  {NoStop}%
\bibitem [{\citenamefont {{Gui}}\ \emph {et~al.}(2020)\citenamefont {{Gui}},
  \citenamefont {{Sun}}, \citenamefont {{Wen}}, \citenamefont {{Tao}},\ and\
  \citenamefont {{Ye}}}]{ReviewGAN}%
  \BibitemOpen
  \bibfield  {author} {\bibinfo {author} {\bibfnamefont {J.}~\bibnamefont
  {{Gui}}}, \bibinfo {author} {\bibfnamefont {Z.}~\bibnamefont {{Sun}}},
  \bibinfo {author} {\bibfnamefont {Y.}~\bibnamefont {{Wen}}}, \bibinfo
  {author} {\bibfnamefont {D.}~\bibnamefont {{Tao}}}, \ and\ \bibinfo {author}
  {\bibfnamefont {J.}~\bibnamefont {{Ye}}},\ }\bibfield  {title} {\enquote
  {\bibinfo {title} {{A Review on Generative Adversarial Networks: Algorithms,
  Theory, and Applications}},}\ }\href@noop {} {\bibfield  {journal} {\bibinfo
  {journal} {arXiv e-prints}\ } (\bibinfo {year} {2020})},\ \Eprint
  {http://arxiv.org/abs/2001.06937} {arXiv:2001.06937 [cs.LG]} \BibitemShut
  {NoStop}%
\bibitem [{\citenamefont {Swendsen}\ and\ \citenamefont
  {Wang}(1987)}]{Swendsen-Wang}%
  \BibitemOpen
  \bibfield  {author} {\bibinfo {author} {\bibfnamefont {R.~H.}\ \bibnamefont
  {Swendsen}}\ and\ \bibinfo {author} {\bibfnamefont {J.-S.}\ \bibnamefont
  {Wang}},\ }\href@noop {} {\bibfield  {journal} {\bibinfo  {journal} {Phys.
  Rev. Lett. 58, 86}\ ,\ \bibinfo {pages} {5}} (\bibinfo {year}
  {1987})}\BibitemShut {NoStop}%
\bibitem [{\citenamefont {Wolff}(1989)}]{Wolf}%
  \BibitemOpen
  \bibfield  {author} {\bibinfo {author} {\bibfnamefont {U.}~\bibnamefont
  {Wolff}},\ }\href@noop {} {\bibfield  {journal} {\bibinfo  {journal} {Phys.
  Rev. Lett.}\ }\textbf {\bibinfo {volume} {361}},\ \bibinfo {pages} {62}
  (\bibinfo {year} {1989})}\BibitemShut {NoStop}%
\bibitem [{\citenamefont {N.~Prokofev}\ and\ \citenamefont
  {Tupitsyn}(1998)}]{worm}%
  \BibitemOpen
  \bibfield  {author} {\bibinfo {author} {\bibfnamefont {B.~S.}\ \bibnamefont
  {N.~Prokofev}}\ and\ \bibinfo {author} {\bibfnamefont {I.}~\bibnamefont
  {Tupitsyn}},\ }\href@noop {} {\bibfield  {journal} {\bibinfo  {journal}
  {Phys. Lett. A 238}\ ,\ \bibinfo {pages} {253}} (\bibinfo {year}
  {1998})}\BibitemShut {NoStop}%
\bibitem [{\citenamefont {H.~G.~Evertz}\ and\ \citenamefont
  {Marcu}(1998)}]{loop1}%
  \BibitemOpen
  \bibfield  {author} {\bibinfo {author} {\bibfnamefont {G.~L.}\ \bibnamefont
  {H.~G.~Evertz}}\ and\ \bibinfo {author} {\bibfnamefont {M.}~\bibnamefont
  {Marcu}},\ }\href@noop {} {\bibfield  {journal} {\bibinfo  {journal} {Phys.
  Rev. Lett. 70}\ ,\ \bibinfo {pages} {875}} (\bibinfo {year}
  {1998})}\BibitemShut {NoStop}%
\bibitem [{\citenamefont {Evertz}(2003)}]{loop2}%
  \BibitemOpen
  \bibfield  {author} {\bibinfo {author} {\bibfnamefont {H.~G.}\ \bibnamefont
  {Evertz}},\ }\href@noop {} {\bibfield  {journal} {\bibinfo  {journal}
  {Advances in Physics}\ ,\ \bibinfo {pages} {52}} (\bibinfo {year}
  {2003})}\BibitemShut {NoStop}%
\bibitem [{\citenamefont {asen}\ and\ \citenamefont
  {Sandvik}(2002)}]{directed-loop1}%
  \BibitemOpen
  \bibfield  {author} {\bibinfo {author} {\bibfnamefont {O.~F.~S.}\
  \bibnamefont {asen}}\ and\ \bibinfo {author} {\bibfnamefont {A.~W.}\
  \bibnamefont {Sandvik}},\ }\href@noop {} {\bibfield  {journal} {\bibinfo
  {journal} {Phys. Rev. E 66,}\ } (\bibinfo {year} {2002})}\BibitemShut
  {NoStop}%
\bibitem [{\citenamefont {F.~Alet}\ and\ \citenamefont
  {Troyer}(2005)}]{directed-loop2}%
  \BibitemOpen
  \bibfield  {author} {\bibinfo {author} {\bibfnamefont {S.~W.}\ \bibnamefont
  {F.~Alet}}\ and\ \bibinfo {author} {\bibfnamefont {M.}~\bibnamefont
  {Troyer}},\ }\href@noop {} {\bibfield  {journal} {\bibinfo  {journal} {Phys.
  Rev. E 71, 036706}\ } (\bibinfo {year} {2005})}\BibitemShut {NoStop}%
\bibitem [{\citenamefont {Dunjko}\ and\ \citenamefont
  {Briegel}(2018)}]{Dunjko_2018}%
  \BibitemOpen
  \bibfield  {author} {\bibinfo {author} {\bibfnamefont {V.}~\bibnamefont
  {Dunjko}}\ and\ \bibinfo {author} {\bibfnamefont {H.~J.}\ \bibnamefont
  {Briegel}},\ }\bibfield  {title} {\enquote {\bibinfo {title} {Machine
  learning {\&} artificial intelligence in the quantum domain: a review of
  recent progress},}\ }\href {\doibase 10.1088/1361-6633/aab406} {\bibfield
  {journal} {\bibinfo  {journal} {Reports on Progress in Physics}\ }\textbf
  {\bibinfo {volume} {81}},\ \bibinfo {pages} {074001} (\bibinfo {year}
  {2018})}\BibitemShut {NoStop}%
\bibitem [{\citenamefont {{Das Sarma}}\ \emph {et~al.}(2019)\citenamefont {{Das
  Sarma}}, \citenamefont {{Deng}},\ and\ \citenamefont
  {{Duan}}}]{PhysicsToday}%
  \BibitemOpen
  \bibfield  {author} {\bibinfo {author} {\bibfnamefont {S.}~\bibnamefont {{Das
  Sarma}}}, \bibinfo {author} {\bibfnamefont {D.-L.}\ \bibnamefont {{Deng}}}, \
  and\ \bibinfo {author} {\bibfnamefont {L.-M.}\ \bibnamefont {{Duan}}},\
  }\bibfield  {title} {\enquote {\bibinfo {title} {{Machine learning meets
  quantum physics}},}\ }\href {\doibase 10.1063/PT.3.4164} {\bibfield
  {journal} {\bibinfo  {journal} {Physics Today}\ }\textbf {\bibinfo {volume}
  {72}},\ \bibinfo {pages} {48} (\bibinfo {year} {2019})},\ \Eprint
  {http://arxiv.org/abs/1903.03516} {arXiv:1903.03516 [physics.pop-ph]}
  \BibitemShut {NoStop}%
\bibitem [{\citenamefont {Mehta}\ \emph {et~al.}(2019)\citenamefont {Mehta},
  \citenamefont {Bukov}, \citenamefont {Wang}, \citenamefont {Day},
  \citenamefont {Richardson}, \citenamefont {Fisher},\ and\ \citenamefont
  {Schwab}}]{MEHTA20191}%
  \BibitemOpen
  \bibfield  {author} {\bibinfo {author} {\bibfnamefont {P.}~\bibnamefont
  {Mehta}}, \bibinfo {author} {\bibfnamefont {M.}~\bibnamefont {Bukov}},
  \bibinfo {author} {\bibfnamefont {C.-H.}\ \bibnamefont {Wang}}, \bibinfo
  {author} {\bibfnamefont {A.~G.}\ \bibnamefont {Day}}, \bibinfo {author}
  {\bibfnamefont {C.}~\bibnamefont {Richardson}}, \bibinfo {author}
  {\bibfnamefont {C.~K.}\ \bibnamefont {Fisher}}, \ and\ \bibinfo {author}
  {\bibfnamefont {D.~J.}\ \bibnamefont {Schwab}},\ }\bibfield  {title}
  {\enquote {\bibinfo {title} {A high-bias, low-variance introduction to
  machine learning for physicists},}\ }\href {\doibase
  https://doi.org/10.1016/j.physrep.2019.03.001} {\bibfield  {journal}
  {\bibinfo  {journal} {Physics Reports}\ }\textbf {\bibinfo {volume} {810}},\
  \bibinfo {pages} {1 } (\bibinfo {year} {2019})}\BibitemShut {NoStop}%
\bibitem [{\citenamefont {Carleo}\ \emph {et~al.}(2019)\citenamefont {Carleo},
  \citenamefont {Cirac}, \citenamefont {Cranmer}, \citenamefont {Daudet},
  \citenamefont {Schuld}, \citenamefont {Tishby}, \citenamefont
  {Vogt-Maranto},\ and\ \citenamefont {Zdeborov\'a}}]{RevModPhys.91.045002}%
  \BibitemOpen
  \bibfield  {author} {\bibinfo {author} {\bibfnamefont {G.}~\bibnamefont
  {Carleo}}, \bibinfo {author} {\bibfnamefont {I.}~\bibnamefont {Cirac}},
  \bibinfo {author} {\bibfnamefont {K.}~\bibnamefont {Cranmer}}, \bibinfo
  {author} {\bibfnamefont {L.}~\bibnamefont {Daudet}}, \bibinfo {author}
  {\bibfnamefont {M.}~\bibnamefont {Schuld}}, \bibinfo {author} {\bibfnamefont
  {N.}~\bibnamefont {Tishby}}, \bibinfo {author} {\bibfnamefont
  {L.}~\bibnamefont {Vogt-Maranto}}, \ and\ \bibinfo {author} {\bibfnamefont
  {L.}~\bibnamefont {Zdeborov\'a}},\ }\bibfield  {title} {\enquote {\bibinfo
  {title} {Machine learning and the physical sciences},}\ }\href {\doibase
  10.1103/RevModPhys.91.045002} {\bibfield  {journal} {\bibinfo  {journal}
  {Rev. Mod. Phys.}\ }\textbf {\bibinfo {volume} {91}},\ \bibinfo {pages}
  {045002} (\bibinfo {year} {2019})}\BibitemShut {NoStop}%
\bibitem [{\citenamefont {Melko}\ \emph {et~al.}(2019)\citenamefont {Melko},
  \citenamefont {Carleo}, \citenamefont {Carrasquilla},\ and\ \citenamefont
  {Cirac}}]{CarleoRBMReview}%
  \BibitemOpen
  \bibfield  {author} {\bibinfo {author} {\bibfnamefont {R.~G.}\ \bibnamefont
  {Melko}}, \bibinfo {author} {\bibfnamefont {G.}~\bibnamefont {Carleo}},
  \bibinfo {author} {\bibfnamefont {J.}~\bibnamefont {Carrasquilla}}, \ and\
  \bibinfo {author} {\bibfnamefont {J.~I.}\ \bibnamefont {Cirac}},\ }\bibfield
  {title} {\enquote {\bibinfo {title} {Restricted boltzmann machines in quantum
  physics},}\ }\href {\doibase 10.1038/s41567-019-0545-1} {\bibfield  {journal}
  {\bibinfo  {journal} {Nature Physics}\ }\textbf {\bibinfo {volume} {15}},\
  \bibinfo {pages} {887} (\bibinfo {year} {2019})}\BibitemShut {NoStop}%
\bibitem [{\citenamefont {Stavros~Efthymiou}\ and\ \citenamefont
  {Melko}(2019)}]{melkosuperresolution}%
  \BibitemOpen
  \bibfield  {author} {\bibinfo {author} {\bibfnamefont {M.~J.~B.}\
  \bibnamefont {Stavros~Efthymiou}}\ and\ \bibinfo {author} {\bibfnamefont
  {R.~G.}\ \bibnamefont {Melko}},\ }\bibfield  {title} {\enquote {\bibinfo
  {title} {Super-resolving the ising model with convolutional neural
  networks},}\ }\href@noop {} {\bibfield  {journal} {\bibinfo  {journal}
  {PHYSICAL REVIEW B 99, 075113}\ } (\bibinfo {year} {2019})}\BibitemShut
  {NoStop}%
\bibitem [{\citenamefont {Liu}\ \emph {et~al.}(2017)\citenamefont {Liu},
  \citenamefont {Qi}, \citenamefont {Meng},\ and\ \citenamefont {Fu}}]{SLMC}%
  \BibitemOpen
  \bibfield  {author} {\bibinfo {author} {\bibfnamefont {J.}~\bibnamefont
  {Liu}}, \bibinfo {author} {\bibfnamefont {Y.}~\bibnamefont {Qi}}, \bibinfo
  {author} {\bibfnamefont {Z.~Y.}\ \bibnamefont {Meng}}, \ and\ \bibinfo
  {author} {\bibfnamefont {L.}~\bibnamefont {Fu}},\ }\bibfield  {title}
  {\enquote {\bibinfo {title} {Self-learning monte carlo method},}\ }\href
  {\doibase 10.1103/PhysRevB.95.041101} {\bibfield  {journal} {\bibinfo
  {journal} {Phys. Rev. B}\ }\textbf {\bibinfo {volume} {95}},\ \bibinfo
  {pages} {041101} (\bibinfo {year} {2017})}\BibitemShut {NoStop}%
\bibitem [{\citenamefont {Liu}\ \emph {et~al.}(2018)\citenamefont {Liu},
  \citenamefont {Xu}, \citenamefont {Qi}, \citenamefont {Sun},\ and\
  \citenamefont {Meng}}]{SLMC2}%
  \BibitemOpen
  \bibfield  {author} {\bibinfo {author} {\bibfnamefont {Z.~H.}\ \bibnamefont
  {Liu}}, \bibinfo {author} {\bibfnamefont {X.~Y.}\ \bibnamefont {Xu}},
  \bibinfo {author} {\bibfnamefont {Y.}~\bibnamefont {Qi}}, \bibinfo {author}
  {\bibfnamefont {K.}~\bibnamefont {Sun}}, \ and\ \bibinfo {author}
  {\bibfnamefont {Z.~Y.}\ \bibnamefont {Meng}},\ }\bibfield  {title} {\enquote
  {\bibinfo {title} {Itinerant quantum critical point with frustration and a
  non-fermi liquid},}\ }\href {\doibase 10.1103/PhysRevB.98.045116} {\bibfield
  {journal} {\bibinfo  {journal} {Phys. Rev. B}\ }\textbf {\bibinfo {volume}
  {98}},\ \bibinfo {pages} {045116} (\bibinfo {year} {2018})}\BibitemShut
  {NoStop}%
\bibitem [{\citenamefont {Xu}\ \emph {et~al.}(2017)\citenamefont {Xu},
  \citenamefont {Qi}, \citenamefont {Liu}, \citenamefont {Fu},\ and\
  \citenamefont {Meng}}]{SLMC3}%
  \BibitemOpen
  \bibfield  {author} {\bibinfo {author} {\bibfnamefont {X.~Y.}\ \bibnamefont
  {Xu}}, \bibinfo {author} {\bibfnamefont {Y.}~\bibnamefont {Qi}}, \bibinfo
  {author} {\bibfnamefont {J.}~\bibnamefont {Liu}}, \bibinfo {author}
  {\bibfnamefont {L.}~\bibnamefont {Fu}}, \ and\ \bibinfo {author}
  {\bibfnamefont {Z.~Y.}\ \bibnamefont {Meng}},\ }\bibfield  {title} {\enquote
  {\bibinfo {title} {Self-learning quantum monte carlo method in interacting
  fermion systems},}\ }\href {\doibase 10.1103/PhysRevB.96.041119} {\bibfield
  {journal} {\bibinfo  {journal} {Phys. Rev. B}\ }\textbf {\bibinfo {volume}
  {96}},\ \bibinfo {pages} {041119} (\bibinfo {year} {2017})}\BibitemShut
  {NoStop}%
\bibitem [{\citenamefont {{Kohshiro}}\ and\ \citenamefont
  {{Nagai}}(2020)}]{RKKYSLMC}%
  \BibitemOpen
  \bibfield  {author} {\bibinfo {author} {\bibfnamefont {H.}~\bibnamefont
  {{Kohshiro}}}\ and\ \bibinfo {author} {\bibfnamefont {Y.}~\bibnamefont
  {{Nagai}}},\ }\bibfield  {title} {\enquote {\bibinfo {title} {{Effective
  Ruderman-Kittel-Kasuya-Yosida-like interaction in diluted double-exchange
  model: self-learning Monte Carlo approach}},}\ }\href@noop {} {\bibfield
  {journal} {\bibinfo  {journal} {arXiv e-prints}\ } (\bibinfo {year}
  {2020})},\ \Eprint {http://arxiv.org/abs/2005.06992} {arXiv:2005.06992
  [cond-mat.dis-nn]} \BibitemShut {NoStop}%
\bibitem [{\citenamefont {{Albergo}}\ \emph {et~al.}(2019)\citenamefont
  {{Albergo}}, \citenamefont {{Kanwar}},\ and\ \citenamefont
  {{Shanahan}}}]{FlowbasedGenModel}%
  \BibitemOpen
  \bibfield  {author} {\bibinfo {author} {\bibfnamefont {M.~S.}\ \bibnamefont
  {{Albergo}}}, \bibinfo {author} {\bibfnamefont {G.}~\bibnamefont {{Kanwar}}},
  \ and\ \bibinfo {author} {\bibfnamefont {P.~E.}\ \bibnamefont {{Shanahan}}},\
  }\bibfield  {title} {\enquote {\bibinfo {title} {{Flow-based generative
  models for Markov chain Monte Carlo in lattice field theory}},}\ }\href
  {\doibase 10.1103/PhysRevD.100.034515} {\bibfield  {journal} {\bibinfo
  {journal} {\prd}\ }\textbf {\bibinfo {volume} {100}},\ \bibinfo {eid}
  {034515} (\bibinfo {year} {2019})},\ \Eprint
  {http://arxiv.org/abs/1904.12072} {arXiv:1904.12072 [hep-lat]} \BibitemShut
  {NoStop}%
\bibitem [{\citenamefont {{Torlai}}\ and\ \citenamefont
  {{Melko}}(2016)}]{MelkoTorlai}%
  \BibitemOpen
  \bibfield  {author} {\bibinfo {author} {\bibfnamefont {G.}~\bibnamefont
  {{Torlai}}}\ and\ \bibinfo {author} {\bibfnamefont {R.~G.}\ \bibnamefont
  {{Melko}}},\ }\bibfield  {title} {\enquote {\bibinfo {title} {{Learning
  thermodynamics with Boltzmann machines}},}\ }\href {\doibase
  10.1103/PhysRevB.94.165134} {\bibfield  {journal} {\bibinfo  {journal}
  {\prb}\ }\textbf {\bibinfo {volume} {94}},\ \bibinfo {eid} {165134} (\bibinfo
  {year} {2016})},\ \Eprint {http://arxiv.org/abs/1606.02718} {arXiv:1606.02718
  [cond-mat.stat-mech]} \BibitemShut {NoStop}%
\bibitem [{\citenamefont {{Morningstar}}\ and\ \citenamefont
  {{Melko}}(2017)}]{2017arXiv170804622M}%
  \BibitemOpen
  \bibfield  {author} {\bibinfo {author} {\bibfnamefont {A.}~\bibnamefont
  {{Morningstar}}}\ and\ \bibinfo {author} {\bibfnamefont {R.~G.}\ \bibnamefont
  {{Melko}}},\ }\bibfield  {title} {\enquote {\bibinfo {title} {{Deep Learning
  the Ising Model Near Criticality}},}\ }\href@noop {} {\bibfield  {journal}
  {\bibinfo  {journal} {arXiv e-prints}\ } (\bibinfo {year} {2017})},\ \Eprint
  {http://arxiv.org/abs/1708.04622} {arXiv:1708.04622 [cond-mat.dis-nn]}
  \BibitemShut {NoStop}%
\bibitem [{\citenamefont {Carleo}\ and\ \citenamefont
  {Troyer}(2017)}]{Carleo602}%
  \BibitemOpen
  \bibfield  {author} {\bibinfo {author} {\bibfnamefont {G.}~\bibnamefont
  {Carleo}}\ and\ \bibinfo {author} {\bibfnamefont {M.}~\bibnamefont
  {Troyer}},\ }\bibfield  {title} {\enquote {\bibinfo {title} {Solving the
  quantum many-body problem with artificial neural networks},}\ }\href
  {\doibase 10.1126/science.aag2302} {\bibfield  {journal} {\bibinfo  {journal}
  {Science}\ }\textbf {\bibinfo {volume} {355}},\ \bibinfo {pages} {602}
  (\bibinfo {year} {2017})}\BibitemShut {NoStop}%
\bibitem [{\citenamefont {Huang}\ and\ \citenamefont
  {Wang}(2017)}]{PhysRevB.95.035105}%
  \BibitemOpen
  \bibfield  {author} {\bibinfo {author} {\bibfnamefont {L.}~\bibnamefont
  {Huang}}\ and\ \bibinfo {author} {\bibfnamefont {L.}~\bibnamefont {Wang}},\
  }\bibfield  {title} {\enquote {\bibinfo {title} {Accelerated monte carlo
  simulations with restricted boltzmann machines},}\ }\href {\doibase
  10.1103/PhysRevB.95.035105} {\bibfield  {journal} {\bibinfo  {journal} {Phys.
  Rev. B}\ }\textbf {\bibinfo {volume} {95}},\ \bibinfo {pages} {035105}
  (\bibinfo {year} {2017})}\BibitemShut {NoStop}%
\bibitem [{\citenamefont {{Zhao}}\ \emph {et~al.}(2019)\citenamefont {{Zhao}},
  \citenamefont {{Kao}}, \citenamefont {{Wu}},\ and\ \citenamefont
  {{Kao}}}]{YingJerIceGeneration}%
  \BibitemOpen
  \bibfield  {author} {\bibinfo {author} {\bibfnamefont {K.-W.}\ \bibnamefont
  {{Zhao}}}, \bibinfo {author} {\bibfnamefont {W.-H.}\ \bibnamefont {{Kao}}},
  \bibinfo {author} {\bibfnamefont {K.-H.}\ \bibnamefont {{Wu}}}, \ and\
  \bibinfo {author} {\bibfnamefont {Y.-J.}\ \bibnamefont {{Kao}}},\ }\bibfield
  {title} {\enquote {\bibinfo {title} {{Generation of ice states through deep
  reinforcement learning}},}\ }\href {\doibase 10.1103/PhysRevE.99.062106}
  {\bibfield  {journal} {\bibinfo  {journal} {\pre}\ }\textbf {\bibinfo
  {volume} {99}},\ \bibinfo {eid} {062106} (\bibinfo {year} {2019})},\ \Eprint
  {http://arxiv.org/abs/1903.04698} {arXiv:1903.04698 [cond-mat.dis-nn]}
  \BibitemShut {NoStop}%
\bibitem [{\citenamefont {{Urban}}\ and\ \citenamefont
  {{Pawlowski}}(2018)}]{ScalarFieldTheoryGAN}%
  \BibitemOpen
  \bibfield  {author} {\bibinfo {author} {\bibfnamefont {J.~M.}\ \bibnamefont
  {{Urban}}}\ and\ \bibinfo {author} {\bibfnamefont {J.~M.}\ \bibnamefont
  {{Pawlowski}}},\ }\bibfield  {title} {\enquote {\bibinfo {title} {{Reducing
  Autocorrelation Times in Lattice Simulations with Generative Adversarial
  Networks}},}\ }\href@noop {} {\bibfield  {journal} {\bibinfo  {journal}
  {arXiv e-prints}\ } (\bibinfo {year} {2018})},\ \Eprint
  {http://arxiv.org/abs/1811.03533} {arXiv:1811.03533 [hep-lat]} \BibitemShut
  {NoStop}%
\bibitem [{\citenamefont {{Mills}}\ and\ \citenamefont
  {{Tamblyn}}(2017)}]{MillsGANIsing}%
  \BibitemOpen
  \bibfield  {author} {\bibinfo {author} {\bibfnamefont {K.}~\bibnamefont
  {{Mills}}}\ and\ \bibinfo {author} {\bibfnamefont {I.}~\bibnamefont
  {{Tamblyn}}},\ }\bibfield  {title} {\enquote {\bibinfo {title} {{Phase space
  sampling and operator confidence with generative adversarial networks}},}\
  }\href@noop {} {\bibfield  {journal} {\bibinfo  {journal} {arXiv e-prints}\ }
  (\bibinfo {year} {2017})},\ \Eprint {http://arxiv.org/abs/1710.08053}
  {arXiv:1710.08053 [cond-mat.stat-mech]} \BibitemShut {NoStop}%
\bibitem [{\citenamefont {Mills}\ \emph {et~al.}()\citenamefont {Mills},
  \citenamefont {Casert},\ and\ \citenamefont {Tamblyn}}]{millsadversarial}%
  \BibitemOpen
  \bibfield  {author} {\bibinfo {author} {\bibfnamefont {K.}~\bibnamefont
  {Mills}}, \bibinfo {author} {\bibfnamefont {C.}~\bibnamefont {Casert}}, \
  and\ \bibinfo {author} {\bibfnamefont {I.}~\bibnamefont {Tamblyn}},\
  }\bibfield  {title} {\enquote {\bibinfo {title} {Adversarial generation of
  mesoscale surfaces from small scale chemical motifs},}\ }\href@noop {} {\
  }\BibitemShut {NoStop}%
\bibitem [{\citenamefont {Zhou}\ \emph {et~al.}(2019)\citenamefont {Zhou},
  \citenamefont {Endr\ifmmode~\mbox{\H{o}}\else \H{o}\fi{}di}, \citenamefont
  {Pang},\ and\ \citenamefont {St\"ocker}}]{PhysRevD.100.011501}%
  \BibitemOpen
  \bibfield  {author} {\bibinfo {author} {\bibfnamefont {K.}~\bibnamefont
  {Zhou}}, \bibinfo {author} {\bibfnamefont {G.}~\bibnamefont
  {Endr\ifmmode~\mbox{\H{o}}\else \H{o}\fi{}di}}, \bibinfo {author}
  {\bibfnamefont {L.-G.}\ \bibnamefont {Pang}}, \ and\ \bibinfo {author}
  {\bibfnamefont {H.}~\bibnamefont {St\"ocker}},\ }\bibfield  {title} {\enquote
  {\bibinfo {title} {Regressive and generative neural networks for scalar field
  theory},}\ }\href {\doibase 10.1103/PhysRevD.100.011501} {\bibfield
  {journal} {\bibinfo  {journal} {Phys. Rev. D}\ }\textbf {\bibinfo {volume}
  {100}},\ \bibinfo {pages} {011501} (\bibinfo {year} {2019})}\BibitemShut
  {NoStop}%
\bibitem [{\citenamefont {{Liu}}\ \emph {et~al.}(2017)\citenamefont {{Liu}},
  \citenamefont {{Rodrigues}},\ and\ \citenamefont {{Cai}}}]{IsingDeepGAN}%
  \BibitemOpen
  \bibfield  {author} {\bibinfo {author} {\bibfnamefont {Z.}~\bibnamefont
  {{Liu}}}, \bibinfo {author} {\bibfnamefont {S.~P.}\ \bibnamefont
  {{Rodrigues}}}, \ and\ \bibinfo {author} {\bibfnamefont {W.}~\bibnamefont
  {{Cai}}},\ }\bibfield  {title} {\enquote {\bibinfo {title} {{Simulating the
  Ising Model with a Deep Convolutional Generative Adversarial Network}},}\
  }\href@noop {} {\bibfield  {journal} {\bibinfo  {journal} {arXiv e-prints}\ }
  (\bibinfo {year} {2017})},\ \Eprint {http://arxiv.org/abs/1710.04987}
  {arXiv:1710.04987 [cond-mat.dis-nn]} \BibitemShut {NoStop}%
\bibitem [{\citenamefont {{Casert}}\ \emph {et~al.}(2020)\citenamefont
  {{Casert}}, \citenamefont {{Mills}}, \citenamefont {{Vieijra}}, \citenamefont
  {{Ryckebusch}},\ and\ \citenamefont {{Tamblyn}}}]{FermiHubbardGAN}%
  \BibitemOpen
  \bibfield  {author} {\bibinfo {author} {\bibfnamefont {C.}~\bibnamefont
  {{Casert}}}, \bibinfo {author} {\bibfnamefont {K.}~\bibnamefont {{Mills}}},
  \bibinfo {author} {\bibfnamefont {T.}~\bibnamefont {{Vieijra}}}, \bibinfo
  {author} {\bibfnamefont {J.}~\bibnamefont {{Ryckebusch}}}, \ and\ \bibinfo
  {author} {\bibfnamefont {I.}~\bibnamefont {{Tamblyn}}},\ }\bibfield  {title}
  {\enquote {\bibinfo {title} {{Optical lattice experiments at unobserved
  conditions and scales through generative adversarial deep learning}},}\
  }\href@noop {} {\bibfield  {journal} {\bibinfo  {journal} {arXiv e-prints}\ }
  (\bibinfo {year} {2020})},\ \Eprint {http://arxiv.org/abs/2002.07055}
  {arXiv:2002.07055 [physics.comp-ph]} \BibitemShut {NoStop}%
\bibitem [{\citenamefont {Cristoforetti}\ \emph {et~al.}(2017)\citenamefont
  {Cristoforetti}, \citenamefont {Jurman}, \citenamefont {Nardelli},\ and\
  \citenamefont {Furlanello}}]{cristoforetti2017towards}%
  \BibitemOpen
  \bibfield  {author} {\bibinfo {author} {\bibfnamefont {M.}~\bibnamefont
  {Cristoforetti}}, \bibinfo {author} {\bibfnamefont {G.}~\bibnamefont
  {Jurman}}, \bibinfo {author} {\bibfnamefont {A.~I.}\ \bibnamefont
  {Nardelli}}, \ and\ \bibinfo {author} {\bibfnamefont {C.}~\bibnamefont
  {Furlanello}},\ }\bibfield  {title} {\enquote {\bibinfo {title} {Towards
  meaningful physics from generative models},}\ }\href@noop {} {\bibfield
  {journal} {\bibinfo  {journal} {arXiv:1705.09524}\ } (\bibinfo {year}
  {2017})}\BibitemShut {NoStop}%
\bibitem [{\citenamefont {Nosarzewski}(2017)}]{1dXYmodel}%
  \BibitemOpen
  \bibfield  {author} {\bibinfo {author} {\bibfnamefont {B.}~\bibnamefont
  {Nosarzewski}},\ }\bibfield  {title} {\enquote {\bibinfo {title}
  {{Variational Autoencoders for Classical Spin Models}},}\ }\href@noop {} {\
  (\bibinfo {year} {2017})}\BibitemShut {NoStop}%
\bibitem [{\citenamefont {I.~Luchnikov}\ and\ \citenamefont
  {Ouerdane}(2019)}]{VAEforPhy}%
  \BibitemOpen
  \bibfield  {author} {\bibinfo {author} {\bibfnamefont {P.~S. S.~F.}\
  \bibnamefont {I.~Luchnikov}, \bibfnamefont {A.~Ryzhov}}\ and\ \bibinfo
  {author} {\bibfnamefont {H.}~\bibnamefont {Ouerdane}},\ }\bibfield  {title}
  {\enquote {\bibinfo {title} {Variational autoencoder reconstruction of
  complex many-body physics.}}\ }\href@noop {} {\bibfield  {journal} {\bibinfo
  {journal} {arxiv 1910.03957}\ } (\bibinfo {year} {2019})}\BibitemShut
  {NoStop}%
\bibitem [{\citenamefont {{Wu}}\ \emph {et~al.}(2019)\citenamefont {{Wu}},
  \citenamefont {{Wang}},\ and\ \citenamefont {{Zhang}}}]{2019PhRvL.122h0602W}%
  \BibitemOpen
  \bibfield  {author} {\bibinfo {author} {\bibfnamefont {D.}~\bibnamefont
  {{Wu}}}, \bibinfo {author} {\bibfnamefont {L.}~\bibnamefont {{Wang}}}, \ and\
  \bibinfo {author} {\bibfnamefont {P.}~\bibnamefont {{Zhang}}},\ }\bibfield
  {title} {\enquote {\bibinfo {title} {{Solving Statistical Mechanics Using
  Variational Autoregressive Networks}},}\ }\href {\doibase
  10.1103/PhysRevLett.122.080602} {\bibfield  {journal} {\bibinfo  {journal}
  {\prl}\ }\textbf {\bibinfo {volume} {122}},\ \bibinfo {eid} {080602}
  (\bibinfo {year} {2019})},\ \Eprint {http://arxiv.org/abs/1809.10606}
  {arXiv:1809.10606 [cond-mat.stat-mech]} \BibitemShut {NoStop}%
\bibitem [{\citenamefont {Sharir}\ \emph {et~al.}(2020)\citenamefont {Sharir},
  \citenamefont {Levine}, \citenamefont {Wies}, \citenamefont {Carleo},\ and\
  \citenamefont {Shashua}}]{PhysRevLett.124.020503}%
  \BibitemOpen
  \bibfield  {author} {\bibinfo {author} {\bibfnamefont {O.}~\bibnamefont
  {Sharir}}, \bibinfo {author} {\bibfnamefont {Y.}~\bibnamefont {Levine}},
  \bibinfo {author} {\bibfnamefont {N.}~\bibnamefont {Wies}}, \bibinfo {author}
  {\bibfnamefont {G.}~\bibnamefont {Carleo}}, \ and\ \bibinfo {author}
  {\bibfnamefont {A.}~\bibnamefont {Shashua}},\ }\bibfield  {title} {\enquote
  {\bibinfo {title} {Deep autoregressive models for the efficient variational
  simulation of many-body quantum systems},}\ }\href {\doibase
  10.1103/PhysRevLett.124.020503} {\bibfield  {journal} {\bibinfo  {journal}
  {Phys. Rev. Lett.}\ }\textbf {\bibinfo {volume} {124}},\ \bibinfo {pages}
  {020503} (\bibinfo {year} {2020})}\BibitemShut {NoStop}%
\bibitem [{\citenamefont {{Ding}}\ and\ \citenamefont
  {{Zhang}}(2020)}]{DingFreeEnergy}%
  \BibitemOpen
  \bibfield  {author} {\bibinfo {author} {\bibfnamefont {X.}~\bibnamefont
  {{Ding}}}\ and\ \bibinfo {author} {\bibfnamefont {B.}~\bibnamefont
  {{Zhang}}},\ }\bibfield  {title} {\enquote {\bibinfo {title} {{Computing
  Absolute Free Energy with Deep Generative Models}},}\ }\href@noop {}
  {\bibfield  {journal} {\bibinfo  {journal} {arXiv e-prints}\ } (\bibinfo
  {year} {2020})},\ \Eprint {http://arxiv.org/abs/2005.00638} {arXiv:2005.00638
  [cond-mat.stat-mech]} \BibitemShut {NoStop}%
\bibitem [{\citenamefont {{Nicoli}}\ \emph {et~al.}(2020)\citenamefont
  {{Nicoli}}, \citenamefont {{Nakajima}}, \citenamefont {{Strodthoff}},
  \citenamefont {{Samek}}, \citenamefont {{M{\"u}ller}},\ and\ \citenamefont
  {{Kessel}}}]{IsingGenerative}%
  \BibitemOpen
  \bibfield  {author} {\bibinfo {author} {\bibfnamefont {K.~A.}\ \bibnamefont
  {{Nicoli}}}, \bibinfo {author} {\bibfnamefont {S.}~\bibnamefont
  {{Nakajima}}}, \bibinfo {author} {\bibfnamefont {N.}~\bibnamefont
  {{Strodthoff}}}, \bibinfo {author} {\bibfnamefont {W.}~\bibnamefont
  {{Samek}}}, \bibinfo {author} {\bibfnamefont {K.-R.}\ \bibnamefont
  {{M{\"u}ller}}}, \ and\ \bibinfo {author} {\bibfnamefont {P.}~\bibnamefont
  {{Kessel}}},\ }\bibfield  {title} {\enquote {\bibinfo {title}
  {{Asymptotically unbiased estimation of physical observables with neural
  samplers}},}\ }\href {\doibase 10.1103/PhysRevE.101.023304} {\bibfield
  {journal} {\bibinfo  {journal} {\pre}\ }\textbf {\bibinfo {volume} {101}},\
  \bibinfo {eid} {023304} (\bibinfo {year} {2020})},\ \Eprint
  {http://arxiv.org/abs/1910.13496} {arXiv:1910.13496 [cond-mat.stat-mech]}
  \BibitemShut {NoStop}%
\bibitem [{\citenamefont {{Mirza}}\ and\ \citenamefont
  {{Osindero}}(2014)}]{ConditionalGANs}%
  \BibitemOpen
  \bibfield  {author} {\bibinfo {author} {\bibfnamefont {M.}~\bibnamefont
  {{Mirza}}}\ and\ \bibinfo {author} {\bibfnamefont {S.}~\bibnamefont
  {{Osindero}}},\ }\bibfield  {title} {\enquote {\bibinfo {title} {{Conditional
  Generative Adversarial Nets}},}\ }\href@noop {} {\bibfield  {journal}
  {\bibinfo  {journal} {arXiv e-prints}\ ,\ \bibinfo {eid} {arXiv:1411.1784}}
  (\bibinfo {year} {2014})},\ \Eprint {http://arxiv.org/abs/1411.1784}
  {arXiv:1411.1784 [cs.LG]} \BibitemShut {NoStop}%
\bibitem [{\citenamefont {Torlai}\ \emph {et~al.}(2019)\citenamefont {Torlai},
  \citenamefont {Timar}, \citenamefont {van Nieuwenburg}, \citenamefont
  {Levine}, \citenamefont {Omran}, \citenamefont {Keesling}, \citenamefont
  {Bernien}, \citenamefont {Greiner}, \citenamefont
  {Vuleti\ifmmode~\acute{c}\else \'{c}\fi{}}, \citenamefont {Lukin},
  \citenamefont {Melko},\ and\ \citenamefont
  {Endres}}]{PhysRevLett.123.230504}%
  \BibitemOpen
  \bibfield  {author} {\bibinfo {author} {\bibfnamefont {G.}~\bibnamefont
  {Torlai}}, \bibinfo {author} {\bibfnamefont {B.}~\bibnamefont {Timar}},
  \bibinfo {author} {\bibfnamefont {E.~P.~L.}\ \bibnamefont {van Nieuwenburg}},
  \bibinfo {author} {\bibfnamefont {H.}~\bibnamefont {Levine}}, \bibinfo
  {author} {\bibfnamefont {A.}~\bibnamefont {Omran}}, \bibinfo {author}
  {\bibfnamefont {A.}~\bibnamefont {Keesling}}, \bibinfo {author}
  {\bibfnamefont {H.}~\bibnamefont {Bernien}}, \bibinfo {author} {\bibfnamefont
  {M.}~\bibnamefont {Greiner}}, \bibinfo {author} {\bibfnamefont
  {V.}~\bibnamefont {Vuleti\ifmmode~\acute{c}\else \'{c}\fi{}}}, \bibinfo
  {author} {\bibfnamefont {M.~D.}\ \bibnamefont {Lukin}}, \bibinfo {author}
  {\bibfnamefont {R.~G.}\ \bibnamefont {Melko}}, \ and\ \bibinfo {author}
  {\bibfnamefont {M.}~\bibnamefont {Endres}},\ }\bibfield  {title} {\enquote
  {\bibinfo {title} {Integrating neural networks with a quantum simulator for
  state reconstruction},}\ }\href {\doibase 10.1103/PhysRevLett.123.230504}
  {\bibfield  {journal} {\bibinfo  {journal} {Phys. Rev. Lett.}\ }\textbf
  {\bibinfo {volume} {123}},\ \bibinfo {pages} {230504} (\bibinfo {year}
  {2019})}\BibitemShut {NoStop}%
\bibitem [{\citenamefont {Shakir~Mohamed}(2017)}]{ImplicitGAN}%
  \BibitemOpen
  \bibfield  {author} {\bibinfo {author} {\bibfnamefont {B.~L.}\ \bibnamefont
  {Shakir~Mohamed}},\ }\bibfield  {title} {\enquote {\bibinfo {title} {Learning
  in implicit generative models.}}\ }\href@noop {} {\bibfield  {journal}
  {\bibinfo  {journal} {arxiv:1610.03483.}\ } (\bibinfo {year}
  {2017})}\BibitemShut {NoStop}%
\bibitem [{\citenamefont {Beach}\ \emph {et~al.}(2018)\citenamefont {Beach},
  \citenamefont {Golubeva},\ and\ \citenamefont {Melko}}]{XYModelSupervised}%
  \BibitemOpen
  \bibfield  {author} {\bibinfo {author} {\bibfnamefont {M.~J.~S.}\
  \bibnamefont {Beach}}, \bibinfo {author} {\bibfnamefont {A.}~\bibnamefont
  {Golubeva}}, \ and\ \bibinfo {author} {\bibfnamefont {R.~G.}\ \bibnamefont
  {Melko}},\ }\bibfield  {title} {\enquote {\bibinfo {title} {Machine learning
  vortices at the kosterlitz-thouless transition},}\ }\href {\doibase
  10.1103/PhysRevB.97.045207} {\bibfield  {journal} {\bibinfo  {journal} {Phys.
  Rev. B}\ }\textbf {\bibinfo {volume} {97}},\ \bibinfo {pages} {045207}
  (\bibinfo {year} {2018})}\BibitemShut {NoStop}%
\bibitem [{\citenamefont {Kosterlitz}\ and\ \citenamefont
  {Thouless}(1973)}]{BKTTransition}%
  \BibitemOpen
  \bibfield  {author} {\bibinfo {author} {\bibfnamefont {J.~M.}\ \bibnamefont
  {Kosterlitz}}\ and\ \bibinfo {author} {\bibfnamefont {D.~J.}\ \bibnamefont
  {Thouless}},\ }\bibfield  {title} {\enquote {\bibinfo {title} {Ordering,
  metastability and phase transitions in two-dimensional systems},}\ }\href
  {http://stacks.iop.org/0022-3719/6/i=7/a=010} {\bibfield  {journal} {\bibinfo
   {journal} {Journal of Physics C: Solid State Physics}\ }\textbf {\bibinfo
  {volume} {6}},\ \bibinfo {pages} {1181} (\bibinfo {year} {1973})}\BibitemShut
  {NoStop}%
\bibitem [{\citenamefont {Berezinskii}(1971)}]{berezinskii}%
  \BibitemOpen
  \bibfield  {author} {\bibinfo {author} {\bibfnamefont {V.}~\bibnamefont
  {Berezinskii}},\ }\bibfield  {title} {\enquote {\bibinfo {title} {Destruction
  of long-range order in one-dimensional and two-dimensional systems having a
  continuous symmetry group i. classical systems},}\ }\href@noop {} {\bibfield
  {journal} {\bibinfo  {journal} {Sov. Phys. JETP}\ }\textbf {\bibinfo {volume}
  {32}},\ \bibinfo {pages} {493} (\bibinfo {year} {1971})}\BibitemShut
  {NoStop}%
\bibitem [{\citenamefont {Berezinskii}(1972)}]{berezinskiiII}%
  \BibitemOpen
  \bibfield  {author} {\bibinfo {author} {\bibfnamefont {V.}~\bibnamefont
  {Berezinskii}},\ }\bibfield  {title} {\enquote {\bibinfo {title} {Destruction
  of long-range order in one-dimensional and two-dimensional systems possessing
  a continuous symmetry group. ii. quantum systems},}\ }\href@noop {}
  {\bibfield  {journal} {\bibinfo  {journal} {Soviet Journal of Experimental
  and Theoretical Physics}\ }\textbf {\bibinfo {volume} {34}},\ \bibinfo
  {pages} {610} (\bibinfo {year} {1972})}\BibitemShut {NoStop}%
\bibitem [{\citenamefont {Kosterlitz}(1974)}]{KosterlitzII}%
  \BibitemOpen
  \bibfield  {author} {\bibinfo {author} {\bibfnamefont {J.~M.}\ \bibnamefont
  {Kosterlitz}},\ }\bibfield  {title} {\enquote {\bibinfo {title} {The critical
  properties of the two-dimensional xy model},}\ }\href
  {http://stacks.iop.org/0022-3719/7/i=6/a=005} {\bibfield  {journal} {\bibinfo
   {journal} {Journal of Physics C: Solid State Physics}\ }\textbf {\bibinfo
  {volume} {7}},\ \bibinfo {pages} {1046} (\bibinfo {year} {1974})}\BibitemShut
  {NoStop}%
\bibitem [{\citenamefont {Rodriguez-Nieva}\ and\ \citenamefont
  {Scheurer}(2019)}]{NatPhysTop}%
  \BibitemOpen
  \bibfield  {author} {\bibinfo {author} {\bibfnamefont {J.~F.}\ \bibnamefont
  {Rodriguez-Nieva}}\ and\ \bibinfo {author} {\bibfnamefont {M.~S.}\
  \bibnamefont {Scheurer}},\ }\bibfield  {title} {\enquote {\bibinfo {title}
  {Identifying topological order through unsupervised machine learning},}\
  }\href {\doibase 10.1038/s41567-019-0512-x} {\bibfield  {journal} {\bibinfo
  {journal} {Nature Physics}\ }\textbf {\bibinfo {volume} {15}},\ \bibinfo
  {pages} {790} (\bibinfo {year} {2019})}\BibitemShut {NoStop}%
\bibitem [{\citenamefont {Long}\ \emph {et~al.}(2020)\citenamefont {Long},
  \citenamefont {Ren},\ and\ \citenamefont {Chen}}]{PhysRevLett.124.185501}%
  \BibitemOpen
  \bibfield  {author} {\bibinfo {author} {\bibfnamefont {Y.}~\bibnamefont
  {Long}}, \bibinfo {author} {\bibfnamefont {J.}~\bibnamefont {Ren}}, \ and\
  \bibinfo {author} {\bibfnamefont {H.}~\bibnamefont {Chen}},\ }\bibfield
  {title} {\enquote {\bibinfo {title} {Unsupervised manifold clustering of
  topological phononics},}\ }\href {\doibase 10.1103/PhysRevLett.124.185501}
  {\bibfield  {journal} {\bibinfo  {journal} {Phys. Rev. Lett.}\ }\textbf
  {\bibinfo {volume} {124}},\ \bibinfo {pages} {185501} (\bibinfo {year}
  {2020})}\BibitemShut {NoStop}%
\bibitem [{\citenamefont {{Che}}\ \emph {et~al.}(2020)\citenamefont {{Che}},
  \citenamefont {{Gneiting}}, \citenamefont {{Liu}},\ and\ \citenamefont
  {{Nori}}}]{2020arXiv200202363C}%
  \BibitemOpen
  \bibfield  {author} {\bibinfo {author} {\bibfnamefont {Y.}~\bibnamefont
  {{Che}}}, \bibinfo {author} {\bibfnamefont {C.}~\bibnamefont {{Gneiting}}},
  \bibinfo {author} {\bibfnamefont {T.}~\bibnamefont {{Liu}}}, \ and\ \bibinfo
  {author} {\bibfnamefont {F.}~\bibnamefont {{Nori}}},\ }\bibfield  {title}
  {\enquote {\bibinfo {title} {{Topological Quantum Phase Transitions Retrieved
  from Manifold Learning}},}\ }\href@noop {} {\bibfield  {journal} {\bibinfo
  {journal} {arXiv e-prints}\ } (\bibinfo {year} {2020})},\ \Eprint
  {http://arxiv.org/abs/2002.02363} {arXiv:2002.02363 [physics.comp-ph]}
  \BibitemShut {NoStop}%
\bibitem [{\citenamefont {{Scheurer}}\ and\ \citenamefont
  {{Slager}}(2020)}]{PRLTopology}%
  \BibitemOpen
  \bibfield  {author} {\bibinfo {author} {\bibfnamefont {M.~S.}\ \bibnamefont
  {{Scheurer}}}\ and\ \bibinfo {author} {\bibfnamefont {R.-J.}\ \bibnamefont
  {{Slager}}},\ }\bibfield  {title} {\enquote {\bibinfo {title} {{Unsupervised
  Machine Learning and Band Topology}},}\ }\href {\doibase
  10.1103/PhysRevLett.124.226401} {\bibfield  {journal} {\bibinfo  {journal}
  {\prl}\ }\textbf {\bibinfo {volume} {124}},\ \bibinfo {eid} {226401}
  (\bibinfo {year} {2020})},\ \Eprint {http://arxiv.org/abs/2001.01711}
  {arXiv:2001.01711 [cond-mat.mes-hall]} \BibitemShut {NoStop}%
\bibitem [{\citenamefont {Sch{\"a}fer}\ and\ \citenamefont
  {L{\"o}rch}(2019)}]{vectordiv}%
  \BibitemOpen
  \bibfield  {author} {\bibinfo {author} {\bibfnamefont {F.}~\bibnamefont
  {Sch{\"a}fer}}\ and\ \bibinfo {author} {\bibfnamefont {N.}~\bibnamefont
  {L{\"o}rch}},\ }\bibfield  {title} {\enquote {\bibinfo {title} {Vector field
  divergence of predictive model output as indication of phase transitions},}\
  }\href@noop {} {\bibfield  {journal} {\bibinfo  {journal} {PHYSICAL REVIEW E
  99, 062107}\ } (\bibinfo {year} {2019})}\BibitemShut {NoStop}%
\bibitem [{\citenamefont {Kashiwa}\ \emph {et~al.}(2019)\citenamefont
  {Kashiwa}, \citenamefont {Kikuchi},\ and\ \citenamefont
  {Tomiya}}]{InWeightsOfNetworks}%
  \BibitemOpen
  \bibfield  {author} {\bibinfo {author} {\bibfnamefont {K.}~\bibnamefont
  {Kashiwa}}, \bibinfo {author} {\bibfnamefont {Y.}~\bibnamefont {Kikuchi}}, \
  and\ \bibinfo {author} {\bibfnamefont {A.}~\bibnamefont {Tomiya}},\
  }\bibfield  {title} {\enquote {\bibinfo {title} {{Phase transition encoded in
  neural network}},}\ }\href {https://doi.org/10.1093/ptep/ptz082} {\bibfield
  {journal} {\bibinfo  {journal} {Progress of Theoretical and Experimental
  Physics}\ }\textbf {\bibinfo {volume} {2019}} (\bibinfo {year} {2019})},\
  \bibinfo {note} {083A04}\BibitemShut {NoStop}%
\bibitem [{\citenamefont {{Tanaka}}\ and\ \citenamefont
  {{Tomiya}}(2017)}]{WeightOfNetwork}%
  \BibitemOpen
  \bibfield  {author} {\bibinfo {author} {\bibfnamefont {A.}~\bibnamefont
  {{Tanaka}}}\ and\ \bibinfo {author} {\bibfnamefont {A.}~\bibnamefont
  {{Tomiya}}},\ }\bibfield  {title} {\enquote {\bibinfo {title} {{Detection of
  Phase Transition via Convolutional Neural Networks}},}\ }\href {\doibase
  10.7566/JPSJ.86.063001} {\bibfield  {journal} {\bibinfo  {journal} {Journal
  of the Physical Society of Japan}\ }\textbf {\bibinfo {volume} {86}},\
  \bibinfo {eid} {063001} (\bibinfo {year} {2017})},\ \Eprint
  {http://arxiv.org/abs/1609.09087} {arXiv:1609.09087 [cond-mat.dis-nn]}
  \BibitemShut {NoStop}%
\bibitem [{\citenamefont {Eliska~Greplova}\ and\ \citenamefont
  {Huber}(2019)}]{topologicalorder}%
  \BibitemOpen
  \bibfield  {author} {\bibinfo {author} {\bibfnamefont {G.~B. F. S. N.~L.}\
  \bibnamefont {Eliska~Greplova}, \bibfnamefont {Agnes~Valenti}}\ and\ \bibinfo
  {author} {\bibfnamefont {S.}~\bibnamefont {Huber}},\ }\bibfield  {title}
  {\enquote {\bibinfo {title} {Unsupervised identification of topological order
  using predictive models},}\ }\href@noop {} {\bibfield  {journal} {\bibinfo
  {journal} {arxiv.org 1910.10124}\ } (\bibinfo {year} {2019})}\BibitemShut
  {NoStop}%
\bibitem [{\citenamefont {Doersch}(2016)}]{VAEtutorial}%
  \BibitemOpen
  \bibfield  {author} {\bibinfo {author} {\bibfnamefont {C.}~\bibnamefont
  {Doersch}},\ }\bibfield  {title} {\enquote {\bibinfo {title} {A tutorial on
  variational autoencoders},}\ }\href@noop {} {\bibfield  {journal} {\bibinfo
  {journal} {arxiv:1606.05908}\ } (\bibinfo {year} {2016})}\BibitemShut
  {NoStop}%
\bibitem [{\citenamefont {Mustafa}\ \emph {et~al.}(2019)\citenamefont
  {Mustafa}, \citenamefont {Bard}, \citenamefont {Bhimji}, \citenamefont
  {Luki{\'c}}, \citenamefont {Al-Rfou},\ and\ \citenamefont
  {Kratochvil}}]{CosmoGAN}%
  \BibitemOpen
  \bibfield  {author} {\bibinfo {author} {\bibfnamefont {M.}~\bibnamefont
  {Mustafa}}, \bibinfo {author} {\bibfnamefont {D.}~\bibnamefont {Bard}},
  \bibinfo {author} {\bibfnamefont {W.}~\bibnamefont {Bhimji}}, \bibinfo
  {author} {\bibfnamefont {Z.}~\bibnamefont {Luki{\'c}}}, \bibinfo {author}
  {\bibfnamefont {R.}~\bibnamefont {Al-Rfou}}, \ and\ \bibinfo {author}
  {\bibfnamefont {J.~M.}\ \bibnamefont {Kratochvil}},\ }\bibfield  {title}
  {\enquote {\bibinfo {title} {Cosmogan: creating high-fidelity weak lensing
  convergence maps using generative adversarial networks},}\ }\href {\doibase
  10.1186/s40668-019-0029-9} {\bibfield  {journal} {\bibinfo  {journal}
  {Computational Astrophysics and Cosmology}\ }\textbf {\bibinfo {volume}
  {6}},\ \bibinfo {pages} {1} (\bibinfo {year} {2019})}\BibitemShut {NoStop}%
\bibitem [{\citenamefont {Tierney}(1992)}]{MH_algo}%
  \BibitemOpen
  \bibfield  {author} {\bibinfo {author} {\bibfnamefont {L.}~\bibnamefont
  {Tierney}},\ }\bibfield  {title} {\enquote {\bibinfo {title} {{Markov Chains
  for Exploring Posterior Distributions}},}\ }\href@noop {} {\bibfield
  {journal} {\bibinfo  {journal} {Annals of Statistics}\ } (\bibinfo {year}
  {1992})}\BibitemShut {NoStop}%
\bibitem [{\citenamefont {Wang}\ and\ \citenamefont {Zhai}(2018)}]{XYPCA}%
  \BibitemOpen
  \bibfield  {author} {\bibinfo {author} {\bibfnamefont {C.}~\bibnamefont
  {Wang}}\ and\ \bibinfo {author} {\bibfnamefont {H.}~\bibnamefont {Zhai}},\
  }\bibfield  {title} {\enquote {\bibinfo {title} {Machine learning of
  frustrated classical spin models (ii): Kernel principal component
  analysis},}\ }\href {\doibase 10.1007/s11467-018-0798-7} {\bibfield
  {journal} {\bibinfo  {journal} {Frontiers of Physics}\ }\textbf {\bibinfo
  {volume} {13}},\ \bibinfo {pages} {130507} (\bibinfo {year}
  {2018})}\BibitemShut {NoStop}%
\bibitem [{\citenamefont {Makhzani}(2019)}]{ImplicitVAE}%
  \BibitemOpen
  \bibfield  {author} {\bibinfo {author} {\bibfnamefont {A.}~\bibnamefont
  {Makhzani}},\ }\bibfield  {title} {\enquote {\bibinfo {title} {Implicit
  autoencoders.}}\ }\href@noop {} {\bibfield  {journal} {\bibinfo  {journal}
  {arxiv 1805.09804}\ } (\bibinfo {year} {2019})}\BibitemShut {NoStop}%
\bibitem [{\citenamefont {Silver}\ \emph {et~al.}(2016)\citenamefont {Silver},
  \citenamefont {Huang}, \citenamefont {Maddison}, \citenamefont {Guez},
  \citenamefont {Sifre}, \citenamefont {van~den Driessche}, \citenamefont
  {Schrittwieser}, \citenamefont {Antonoglou}, \citenamefont {Panneershelvam},
  \citenamefont {Lanctot}, \citenamefont {Dieleman}, \citenamefont {Grewe},
  \citenamefont {Nham}, \citenamefont {Kalchbrenner}, \citenamefont
  {Sutskever}, \citenamefont {Lillicrap}, \citenamefont {Leach}, \citenamefont
  {Kavukcuoglu}, \citenamefont {Graepel},\ and\ \citenamefont
  {Hassabis}}]{Alphago}%
  \BibitemOpen
  \bibfield  {author} {\bibinfo {author} {\bibfnamefont {D.}~\bibnamefont
  {Silver}}, \bibinfo {author} {\bibfnamefont {A.}~\bibnamefont {Huang}},
  \bibinfo {author} {\bibfnamefont {C.~J.}\ \bibnamefont {Maddison}}, \bibinfo
  {author} {\bibfnamefont {A.}~\bibnamefont {Guez}}, \bibinfo {author}
  {\bibfnamefont {L.}~\bibnamefont {Sifre}}, \bibinfo {author} {\bibfnamefont
  {G.}~\bibnamefont {van~den Driessche}}, \bibinfo {author} {\bibfnamefont
  {J.}~\bibnamefont {Schrittwieser}}, \bibinfo {author} {\bibfnamefont
  {I.}~\bibnamefont {Antonoglou}}, \bibinfo {author} {\bibfnamefont
  {V.}~\bibnamefont {Panneershelvam}}, \bibinfo {author} {\bibfnamefont
  {M.}~\bibnamefont {Lanctot}}, \bibinfo {author} {\bibfnamefont
  {S.}~\bibnamefont {Dieleman}}, \bibinfo {author} {\bibfnamefont
  {D.}~\bibnamefont {Grewe}}, \bibinfo {author} {\bibfnamefont
  {J.}~\bibnamefont {Nham}}, \bibinfo {author} {\bibfnamefont {N.}~\bibnamefont
  {Kalchbrenner}}, \bibinfo {author} {\bibfnamefont {I.}~\bibnamefont
  {Sutskever}}, \bibinfo {author} {\bibfnamefont {T.}~\bibnamefont
  {Lillicrap}}, \bibinfo {author} {\bibfnamefont {M.}~\bibnamefont {Leach}},
  \bibinfo {author} {\bibfnamefont {K.}~\bibnamefont {Kavukcuoglu}}, \bibinfo
  {author} {\bibfnamefont {T.}~\bibnamefont {Graepel}}, \ and\ \bibinfo
  {author} {\bibfnamefont {D.}~\bibnamefont {Hassabis}},\ }\bibfield  {title}
  {\enquote {\bibinfo {title} {Mastering the game of go with deep neural
  networks and tree search},}\ }\href {\doibase 10.1038/nature16961} {\bibfield
   {journal} {\bibinfo  {journal} {Nature}\ }\textbf {\bibinfo {volume}
  {529}},\ \bibinfo {pages} {484} (\bibinfo {year} {2016})}\BibitemShut
  {NoStop}%
\bibitem [{\citenamefont {Adji B.~Dieng}\ and\ \citenamefont
  {Titsias}(2019)}]{PresGAN}%
  \BibitemOpen
  \bibfield  {author} {\bibinfo {author} {\bibfnamefont {D.~M.~B.}\
  \bibnamefont {Adji B.~Dieng}, \bibfnamefont {Francisco J. R.~Ruiz}}\ and\
  \bibinfo {author} {\bibfnamefont {M.~K.}\ \bibnamefont {Titsias}},\
  }\bibfield  {title} {\enquote {\bibinfo {title} {Prescribed generative
  adversarial networks},}\ }\href@noop {} {\bibfield  {journal} {\bibinfo
  {journal} {arxiv 1910.04302}\ } (\bibinfo {year} {2019})}\BibitemShut
  {NoStop}%
\bibitem [{\citenamefont {{Chen}}\ \emph {et~al.}(2016)\citenamefont {{Chen}},
  \citenamefont {{Duan}}, \citenamefont {{Houthooft}}, \citenamefont
  {{Schulman}}, \citenamefont {{Sutskever}},\ and\ \citenamefont
  {{Abbeel}}}]{Infogan}%
  \BibitemOpen
  \bibfield  {author} {\bibinfo {author} {\bibfnamefont {X.}~\bibnamefont
  {{Chen}}}, \bibinfo {author} {\bibfnamefont {Y.}~\bibnamefont {{Duan}}},
  \bibinfo {author} {\bibfnamefont {R.}~\bibnamefont {{Houthooft}}}, \bibinfo
  {author} {\bibfnamefont {J.}~\bibnamefont {{Schulman}}}, \bibinfo {author}
  {\bibfnamefont {I.}~\bibnamefont {{Sutskever}}}, \ and\ \bibinfo {author}
  {\bibfnamefont {P.}~\bibnamefont {{Abbeel}}},\ }\bibfield  {title} {\enquote
  {\bibinfo {title} {{InfoGAN: Interpretable Representation Learning by
  Information Maximizing Generative Adversarial Nets}},}\ }\href@noop {}
  {\bibfield  {journal} {\bibinfo  {journal} {arXiv e-prints}\ } (\bibinfo
  {year} {2016})},\ \Eprint {http://arxiv.org/abs/1606.03657} {arXiv:1606.03657
  [cs.LG]} \BibitemShut {NoStop}%
\bibitem [{\citenamefont {Quan}\ and\ \citenamefont
  {Cucchietti}(2009)}]{PhysRevE.79.031101}%
  \BibitemOpen
  \bibfield  {author} {\bibinfo {author} {\bibfnamefont {H.~T.}\ \bibnamefont
  {Quan}}\ and\ \bibinfo {author} {\bibfnamefont {F.~M.}\ \bibnamefont
  {Cucchietti}},\ }\bibfield  {title} {\enquote {\bibinfo {title} {Quantum
  fidelity and thermal phase transitions},}\ }\href {\doibase
  10.1103/PhysRevE.79.031101} {\bibfield  {journal} {\bibinfo  {journal} {Phys.
  Rev. E}\ }\textbf {\bibinfo {volume} {79}},\ \bibinfo {pages} {031101}
  (\bibinfo {year} {2009})}\BibitemShut {NoStop}%
\bibitem [{\citenamefont {Komura}\ and\ \citenamefont
  {Okabe}(2012)}]{CriticalTemperatureBKT}%
  \BibitemOpen
  \bibfield  {author} {\bibinfo {author} {\bibfnamefont {Y.}~\bibnamefont
  {Komura}}\ and\ \bibinfo {author} {\bibfnamefont {Y.}~\bibnamefont {Okabe}},\
  }\bibfield  {title} {\enquote {\bibinfo {title} {Large-scale monte carlo
  simulation of two-dimensional classical xy model using multiple gpus},}\
  }\href {\doibase 10.1143/JPSJ.81.113001} {\bibfield  {journal} {\bibinfo
  {journal} {Journal of the Physical Society of Japan}\ }\textbf {\bibinfo
  {volume} {81}},\ \bibinfo {pages} {113001} (\bibinfo {year} {2012})},\
  \Eprint {http://arxiv.org/abs/https://doi.org/10.1143/JPSJ.81.113001}
  {https://doi.org/10.1143/JPSJ.81.113001} \BibitemShut {NoStop}%
\bibitem [{\citenamefont {{Wang}}\ \emph {et~al.}(2020)\citenamefont {{Wang}},
  \citenamefont {{Jiang}}, \citenamefont {{He}},\ and\ \citenamefont
  {{Zhou}}}]{XYModelVAN}%
  \BibitemOpen
  \bibfield  {author} {\bibinfo {author} {\bibfnamefont {L.}~\bibnamefont
  {{Wang}}}, \bibinfo {author} {\bibfnamefont {Y.}~\bibnamefont {{Jiang}}},
  \bibinfo {author} {\bibfnamefont {L.}~\bibnamefont {{He}}}, \ and\ \bibinfo
  {author} {\bibfnamefont {K.}~\bibnamefont {{Zhou}}},\ }\bibfield  {title}
  {\enquote {\bibinfo {title} {{Recognizing the topological phase transition by
  Variational Autoregressive Networks}},}\ }\href@noop {} {\bibfield  {journal}
  {\bibinfo  {journal} {arXiv e-prints}\ } (\bibinfo {year} {2020})},\ \Eprint
  {http://arxiv.org/abs/2005.04857} {arXiv:2005.04857 [cond-mat.dis-nn]}
  \BibitemShut {NoStop}%
\end{thebibliography}

\providecommand{\noopsort}[1]{}\providecommand{\singleletter}[1]{#1}%

\end{document}